\documentclass[twocolumn]{aastex702}

\received{2026 September 9}
\revised{2026 November 10}
\accepted{2026 November 11}
\makeatletter\def\@submitted{Accepted for publication in ApJ}\makeatother
\shorttitle{Variable millimeter excess in NGC 985}
\shortauthors{Doi \& Inoue}

\newcommand{\target}{NGC 985}

\usepackage{placeins}

\begin{document}

\title{A Variable Compact Millimeter Excess in the Type-1 Seyfert Galaxy NGC 985: Implications for a Black-Hole Corona}

\author[0000-0003-4384-9568]{Akihiro Doi}
\affiliation{The Institute of Space and Astronautical Science, Japan Aerospace Exploration Agency, 3-1-1 Yoshinodai, Chuo-ku, Sagamihara, Kanagawa 252-5210, Japan}
\affiliation{Department of Space and Astronautical Science, The Graduate University for Advanced Studies (SOKENDAI), 3-1-1 Yoshinodai, Chuo-ku, Sagamihara, Kanagawa 252-5210, Japan}
\email{doi.akihiro@jaxa.jp}

\author[0000-0002-7272-1136]{Yoshiyuki Inoue}
\affiliation{College of Systems Engineering and Science, Shibaura Institute of Technology, 307 Fukasaku, Minuma-ku, Saitama-shi, Saitama 337-8570, Japan}
\affiliation{RIKEN Center for Interdisciplinary Theoretical and Mathematical Sciences (iTHEMS), 2-1 Hirosawa, Wako 351-0198, Saitama, Japan}
\affiliation{Kavli Institute for the Physics and Mathematics of the Universe (WPI), UTIAS, The University of Tokyo, Kashiwa, Chiba 277-8583, Japan}
\email{yinoue@shibaura-it.ac.jp}

\begin{abstract}
Compact millimeter emission from radio-quiet active galactic nuclei has been proposed as a probe of magnetized accretion-disk coronae. We present three epochs of ALMA observations of the type-1 Seyfert galaxy
NGC~985, together with archival post-upgrade VLA data presented here for
the first time, to investigate its compact millimeter excess.
The millimeter flux density decreased strongly between the first two ALMA
epochs separated by approximately one month, while the component was
detected again at an intermediate level in the third epoch.  This variability
demonstrates an intrinsically variable compact component and implies a
model-independent size limit of $\lesssim0.024$~pc
($\simeq1.1\times10^3R_{\rm S}$).  Spectral decomposition
requires a compact, self-absorbed synchrotron component to reproduce the
millimeter-wave data.  A hybrid thermal/nonthermal coronal
model reproduces the radio--millimeter spectra, with the two broadly sampled
ALMA epochs favoring different nonthermal electron distributions. The estimated radiative cooling times are comparable to or shorter than the light-crossing time, implying continuous injection or reacceleration of nonthermal electrons; a change in particle injection is the most natural driver of the observed spectral changes.  We
conclude that NGC~985 hosts a recurrent and variable compact millimeter
synchrotron component.  A magnetized corona provides a
plausible origin, although the present data do not uniquely distinguish a
coronal origin from alternatives such as a jet base.
\end{abstract}

\section{Introduction}
\label{sec:introduction}

The origin of radio emission from radio-quiet active galactic nuclei
(AGNs) remains less well understood than that from radio-loud AGNs \citep{Panessa:2019}.
A long-standing observational clue is that some radio-quiet
quasars and Seyfert nuclei exhibit flat or inverted high-frequency radio
spectra, or an excess above the extrapolation of the steep-spectrum
centimeter component
\citep{Antonucci:1988,Barvainis:1996,Doi:2005,Doi:2005a,Doi:2011}.
Such high-frequency components are difficult to attribute solely to the extended synchrotron emission associated with star formation in the host galaxy \citep{Condon:1992} and instead point to a more compact nuclear emitter. At millimeter wavelengths, where steep-spectrum extended synchrotron emission becomes less important, these compact components are more readily isolated, although thermal free--free and dust emission may still contribute. Millimeter observations therefore provide a particularly useful probe of compact emission associated with the innermost regions of radio-quiet AGNs.

One possible origin of such compact emission is synchrotron radiation from the magnetized plasma associated with the X-ray-emitting corona. \citet{Laor:2008} noted that the 5-GHz-to-X-ray luminosity relation of radio-quiet quasars closely resembles the G\"udel--Benz relation for coronally active stars \citep{Guedel:1993}. Because synchrotron self-absorption suppresses emission from compact coronal regions at lower radio frequencies, these signatures are expected to become more accessible toward the millimeter and submillimeter bands. In particular, a hybrid thermal/nonthermal electron population can produce detectable synchrotron emission from an AGN corona at these frequencies \citep{Inoue:2014}. Observationally, high-frequency or millimeter excesses above the extrapolations of their lower-frequency radio spectra have been found in several radio-quiet AGNs \citep{Behar:2015}. Recent high-angular-resolution ALMA studies have found correlations between nuclear millimeter and X-ray luminosities \citep{Kawamuro:2022,Ricci:2023,Venselaar:submitted}. These results support a close connection between compact millimeter emission and accretion activity, although they do not uniquely establish that the millimeter-emitting plasma is identical to the compact X-ray-producing corona.

The nonthermal activity of AGN coronae is also relevant to high-energy neutrino production. IceCube has reported evidence for neutrino emission from NGC~1068 and, more recently, from a population of X-ray-bright AGNs \citep{IceCube:2020,IceCube:2022,IceCube:2026}. In NGC~1068, the high neutrino-to-gamma-ray flux ratio favors a compact, gamma-ray-opaque production site, for which an AGN corona is a candidate; millimeter spectra constrain the nonthermal electrons and magnetic fields entering such models \citep{Inoue:2020}.

Time-domain observations provide an additional and largely independent test of the compact nature of the millimeter emission. Significant millimeter variability on day-to-month timescales has been reported in several radio-quiet Seyfert nuclei, including NGC~7469, MCG+08$-$11$-$11, GRS~1734$-$292, IC~4329A, and NGC~1566 \citep{Baldi:2015,Behar:2020,Petrucci:2023,Michiyama:2024,Shablovinskaya:2024,Jana:2025,Venselaar:submitted}. Such variability constrains the source size through causality arguments that are independent of detailed spectral decomposition, thereby offering strong evidence for a compact nuclear origin. An independent constraint comes from gravitational microlensing of the radio-quiet quasar RXJ1131$-$1231, which limits the half-light radius of its millimeter-emitting region to $\lesssim 50$~AU \citep{Rybak:2025}. At the same time, simultaneous or quasi-simultaneous millimeter and X-ray monitoring has not yet revealed a universal short-timescale correspondence between the two bands \citep{Behar:2020,Petrucci:2023,Shablovinskaya:2024}. Multi-epoch measurements of both the strength and spectral shape of the millimeter component are therefore important for determining how this compact synchrotron-emitting plasma is related to the X-ray corona and other nuclear structures.

NGC~985 is a particularly well-suited target in this context. It is a type-1 Seyfert galaxy and was one of the clearest early examples of a high-frequency excess in a radio-quiet luminous AGN \citep{Antonucci:1988,Barvainis:1996}. Subsequent quasi-simultaneous radio observations confirmed the spectral rise toward a compact high-frequency component at tens of GHz \citep{Doi:2016}. Atacama Large Millimeter/submillimeter Array (ALMA) observations later demonstrated that the excess extends into the millimeter regime and that its spectrum can be described by synchrotron emission from a hybrid thermal/nonthermal corona \citep{Inoue:2018}. These observations established NGC~985 as one of the clearest examples of a compact millimeter excess in a radio-quiet AGN and an important candidate for emission associated with a supermassive black hole corona. Its temporal behavior, however, remains largely unexplored: it is unknown whether the millimeter excess represents a relatively stable spectral feature or a compact component that varies between distinct emission states.

In this paper, we investigate the compact millimeter component in NGC~985
using new ALMA observations, together with the
published ALMA data \citep{Inoue:2018} and archival post-upgrade Karl G. Jansky Very Large Array (VLA)
observations.  The latter provide new centimeter-wave constraints relative to
the pre-upgrade VLA data used in \citet{Doi:2016}.  The resulting
multi-epoch ALMA data allow us to examine the temporal and spectral
behavior of the millimeter component within a hybrid-corona synchrotron framework, and to assess the constraints that its variability
places on the nature and size of the compact emitting region.

Throughout this paper, we adopt a $\Lambda$CDM cosmology with
$H_0=70.5$ km s$^{-1}$ Mpc$^{-1}$, $\Omega_{\rm M}=0.27$, and
$\Omega_{\Lambda}=0.73$.  For the redshift
$z=0.043143\pm0.000073$ \citep{Fisher:1995}, the luminosity distance is
186 Mpc and the angular-size distance is 171 Mpc; $1''$ corresponds to a
projected linear scale of 831 pc.

\section{Observations and Data Reduction}\label{section:observation}
\subsection{ALMA Observations and Imaging}
\label{section:ALMA_observation}

We conducted continuum observations of the NGC~985 nucleus with
ALMA during Cycle~3 using the Band~3 and Band~4 receivers
(project ID: 2015.1.01516.S, PI = A. Doi).
Four spectral windows (SPWs) with a bandwidth of 2~GHz were centered at sky frequencies of 85.005, 87.005, 97.068, and 99.005~GHz in Band~3
and at 147.995, 149.932, 159.995, and 161.995~GHz in Band~4.
The data were correlated in the Time Division Mode (TDM) with 128 channels per SPW.
The first observation was made on 2016 March 8 (Epoch-1a) in
Band~3 and Band~4; baselines ranged from 15.1~m to 460~m. Although the first
Band~4 execution passed the initial QA0 assessment, the reduced products
subsequently failed to meet the quality standards for delivery. The
scheduling block was therefore returned to the observing queue for
additional observations, which were carried out on 2016 April 7 (Epoch-1b),
with baselines ranging from 15.1~m to 391.1~m. Both Band~4 executions were
ultimately retained as usable data. In this work, however, we analyze them
separately to investigate possible variability between them.
J0238$+$1636 and J0241$-$0815 were observed as the flux and phase calibrators, respectively.

The data were calibrated using the Common Astronomy Software Applications
(CASA; \citealt{McMullin:2007}) version 4.7.0.
We reduced the raw data using pipeline scripts provided by the ALMA Regional Center (ARC).
In the pipeline, flux scaling factors were determined with reference to
J0238$+$1636, whose flux densities were monitored by the ALMA Observatory.
We queried the ALMA calibrator catalog and found that at 103.5~GHz this quasar remained stable ($\sim2.2$~Jy) around Epoch-1a and decreased gradually
(from $\sim2.2$~Jy to $\sim1.7$~Jy) around Epoch-1b.
The zenith precipitable water vapor was 2.37~mm, 2.85~mm, and 2.81~mm during
the observations of Band~3 at Epoch-1a, Band~4 at Epoch-1a, and Band~4 at
Epoch-1b, respectively.
We found no appreciable line emission within the observed NGC~985 frequency ranges.
The calibrated uv-data were exported for four individual SPWs and imaged and deconvolved with CLEAN using natural weighting.
This procedure yielded four continuum images for Band~3 and eight images
for Band~4.
Point-like emission was detected in all images.
Flux densities were measured by fitting a Gaussian with the CASA task
\texttt{imfit}.
The results are listed in Table~\ref{table:observation}.  
The flux-density uncertainties in Table~\ref{table:observation} combine the formal Gaussian-fit errors returned by \texttt{imfit} with 5\% absolute flux-calibration uncertainty for Bands~3 and 4, and 10\% for Band~6, added in quadrature. Within each observing epoch and ALMA receiver band, the calibration term is common to the SPWs and is treated as a correlated error in the joint spectral fit rather than as independent noise.

\begin{table*}
\caption{Results of ALMA Observations\label{table:observation}}
\begin{center}
\begin{footnotesize}
\begin{tabular}{lcccccccc}
\hline
\hline
Date	&	$\nu$	&	$F_\nu$			&	$\sigma$	&	$\theta_\mathrm{maj} \times \theta_\mathrm{min}$			&	$PA$	&	$F_\nu^\mathrm{re}$			&	$\theta_\mathrm{maj}^\mathrm{re} \times \theta_\mathrm{min}^\mathrm{re}$	\\
(yyyy-mm-dd)	&	(GHz)	&	(mJy)			&	(mJy/beam)	&	(arcsec$\times$arcsec)			&	(deg)	&	(mJy)			&		\\
(1)	&	(2)	&	(3)			&	(4)	&	(5)			&	(6)	&	(7)			&	(8)	\\
\hline
2016-03-08	&	85.0	& $	3.07 	\pm 0.18 	$ & 	0.11 	& $	2.47	\times	1.87	$ & $	78.1	$ & $	3.42	\pm	0.18	$ & 	$2.18 \times 2.18$	\\
(Epoch-1a)	&	87.0	& $	2.93 	\pm 0.17 	$ & 	0.10 	& $	2.40	\times	1.85	$ & $	77.5	$ & $	3.18	\pm	0.19	$ & 		\\
	&	97.1	& $	2.80 	\pm 0.17 	$ & 	0.10 	& $	2.14	\times	1.76	$ & $	81.6	$ & $	3.04	\pm	0.23	$ & 		\\
	&	99.0	& $	2.83 	\pm 0.17 	$ & 	0.10 	& $	2.09	\times	1.70	$ & $	82.7	$ & $	3.16	\pm	0.20	$ & 		\\
	&	148.0	& $	2.24 	\pm 0.15 	$ & 	0.11 	& $	1.53	\times	1.16	$ & $	86.7	$ & $	2.35	\pm	0.16	$ & 		\\
	&	149.9	& $	2.20 	\pm 0.15 	$ & 	0.11 	& $	1.53	\times	1.14	$ & $	87.5	$ & $	2.13	\pm	0.11	$ & 		\\
	&	160.0	& $	2.29 	\pm 0.16 	$ & 	0.10 	& $	1.46	\times	1.06	$ & $	86.6	$ & $	2.34	\pm	0.15	$ & 		\\
	&	162.0	& $	2.07 	\pm 0.16 	$ & 	0.12 	& $	1.42	\times	1.05	$ & $	86.0	$ & $	2.16	\pm	0.17	$ & 		\\
2016-04-07	&	148.0	& $	0.92 	\pm 0.08 	$ & 	0.06 	& $	1.44	\times	1.41	$ & $	-46.3	$ & $	1.13	\pm	0.09	$ & 	$2.18 \times 2.18$	\\
(Epoch-1b)	&	149.9	& $	0.81 	\pm 0.07 	$ & 	0.06 	& $	1.49	\times	1.38	$ & $	4.0	$ & $	1.10	\pm	0.08	$ & 		\\
	&	160.0	& $	0.80 	\pm 0.08 	$ & 	0.06 	& $	1.34	\times	1.32	$ & $	-10.5	$ & $	0.89	\pm	0.07	$ & 		\\
	&	162.0	& $	0.66 	\pm 0.09 	$ & 	0.08 	& $	1.40	\times	1.28	$ & $	28.3	$ & $	0.86	\pm	0.09	$ & 		\\
2017-08-18	&	90.5	& $	1.72 	\pm 0.11 	$ & 	0.06 	& $	0.25	\times	0.22	$ & $	0.0	$ & $	1.92	\pm	0.11	$ & 	$0.27 \times 0.27$	\\
(Epoch-2)	&	92.4	& $	1.92 	\pm 0.12 	$ & 	0.05 	& $	0.25	\times	0.21	$ & $	98.9	$ & $	2.03	\pm	0.12	$ & 		\\
	&	102.5	& $	1.64 	\pm 0.11 	$ & 	0.06 	& $	0.23	\times	0.19	$ & $	84.7	$ & $	1.74	\pm	0.09	$ & 		\\
	&	104.5	& $	1.59 	\pm 0.11 	$ & 	0.07 	& $	0.22	\times	0.19	$ & $	101.2	$ & $	1.85	\pm	0.12	$ & 		\\
	&	138.1	& $	1.71 	\pm 0.12 	$ & 	0.06 	& $	0.17	\times	0.14	$ & $	100.0	$ & $	1.94	\pm	0.11	$ & 		\\
	&	139.9	& $	1.54 	\pm 0.11 	$ & 	0.05 	& $	0.18	\times	0.15	$ & $	83.2	$ & $	1.81	\pm	0.10	$ & 		\\
	&	150.0	& $	1.44 	\pm 0.11 	$ & 	0.05 	& $	0.16	\times	0.13	$ & $	99.0	$ & $	1.68	\pm	0.10	$ & 		\\
	&	151.8	& $	1.57 	\pm 0.11 	$ & 	0.06 	& $	0.16	\times	0.13	$ & $	100.3	$ & $	1.69	\pm	0.11	$ & 		\\
	&	213.0	& $	1.35 	\pm 0.15 	$ & 	0.04 	& $	0.12	\times	0.10	$ & $	84.4	$ & $	1.76	\pm 0.18	$ & 		\\
	&	215.0	& $	1.27 	\pm 0.14 	$ & 	0.04 	& $	0.12	\times	0.09	$ & $	83.2	$ & $	1.65	\pm 0.20	$ & 		\\
	&	229.2	& $	1.26 	\pm 0.14 	$ & 	0.04 	& $	0.11	\times	0.09	$ & $	85.6	$ & $	1.77	\pm 0.21	$ & 		\\
	&	231.0	& $	1.22 	\pm 0.13 	$ & 	0.04 	& $	0.11	\times	0.09	$ & $	86.4	$ & $	1.76	\pm 0.18	$ & 		\\			
\hline
\end{tabular}
\end{footnotesize}
\end{center}
\begin{flushleft}
\begin{footnotesize}
Col.~(1) observation date; 
Col.~(2) center frequency; 
Col.~(3) flux density;   
Col.~(4) image r.m.s.~noise on blank sky; 
Col.~(5)--(6) synthesized beam sizes in major axis, minor axis, and position angle of major axis, respectively.  
Col.~(7) flux density in the image using a restoring beam; 
Col.~(8) restoring beam size. The flux-density errors combine formal Gaussian-fit uncertainties with 5\% absolute calibration uncertainty for Bands~3/4 and 10\% for Band~6.  
\end{footnotesize}
\end{flushleft}
\end{table*}

We also produced beam-matched images at a common resolution across SPWs and epochs, using uvtaper and beam restoration, to avoid
frequency-dependent resolution effects in photometry.
The target resolution was $2.18''$, which was the geometric mean of the major
and minor axes of the synthesized beam at SPW0 in Band~3 with Briggs
weighting \citep{Briggs:1995} and robust$=0.5$.
With the uvtaper process alone, the resultant synthesized beams varied slightly because of differences in the original uv coverage.
Therefore, the images were subsequently restored with a circular beam of $2.18''$.
We applied uvtapering and beam restoration to the Briggs-weighted data in
Band~3 and the natural-weighted data in Band~4.
The \texttt{imfit} results are listed in Table~\ref{table:observation}.

We also utilized the previously calibrated and published Cycle~4 data
\citep{Inoue:2018} (project ID: 2016.1.01140.S, PI = Y. Inoue),
obtained on 2017 August~18 (hereafter Epoch-2).
These data yielded twelve images of the NGC~985 nucleus in Bands~3, 4, and 6, made with natural weighting.
We use these flux-density measurements in the present study; see
\citet{Inoue:2018} for details of the data and data-reduction procedures.

We reanalyzed these calibrated datasets to obtain beam-matched images
at different frequencies.
Imaging followed the Cycle~3 procedure above.
The images were restored with a circular beam of $0.27''$, which corresponds to the synthesized beam size at SPW0 in Band~3 with Briggs weighting and robust$=0.5$.
The \texttt{imfit} results are listed in Table~\ref{table:observation}.

The final usable on-source integration times in the calibrated data were
2.02 and 4.70~min in Bands~3 and 4 of Epoch-1a, respectively;
9.41~min in Band~4 of Epoch-1b; and 5.14, 5.14, and 18.14~min in
Bands~3, 4, and 6 of Epoch-2, respectively. These durations count target
integrations with at least one unflagged visibility and do not scale
individual integrations by their fractional flagging.

In addition, we obtained deeper continuum images by combining the four SPWs in each of Bands~4 and 6.  To search for extended emission components that could increase the measured flux density of the nucleus in photometry
with finite angular resolution, we adopted multi-term multi-frequency
synthesis (MT-MFS; \citealt{Rau:2011}) as the deconvolution algorithm in the
imaging procedure.  The resulting image r.m.s. noise levels were
0.023 and 0.017~mJy~beam$^{-1}$ in Bands~4 and 6, respectively.

\subsection{Intrahour Variability Analysis}
\label{section:intrahour_observation}

We examined intrahour variability during the ALMA observations by constructing a series of consecutive continuum images from the calibrated
visibilities.  The four SPWs were combined, and images were synthesized
with an integration time of $t_{\rm integ}$ at intervals of
$t_{\rm integ}/2$, so that adjacent segments overlapped by half their duration.

The value of $t_{\rm integ}$ for each frequency band and epoch was chosen
to achieve a signal-to-noise ratio (S/N) of approximately 20, allowing
variability with an amplitude of roughly 15\% to be detected at the
$3\sigma$ level.  The adopted values of $t_{\rm integ}$ ranged from 16 to 300~s.  
Phase-calibrator scans occasionally fell within a time segment,
reducing the effective integration time on the target and potentially
degrading the corresponding image.  We therefore excluded flux-density
measurements with S/N below 14.

\subsection{VLA Observations and Imaging}
\label{section:VLAdata}

In this study, we analyzed all archival post-upgrade VLA observations of NGC~985 obtained since 2013 and retrieved from the Science Data
Archive of the National Radio Astronomy Observatory (NRAO).
Except for project 13A-281, these VLA observations were conducted between
2016 March and 2017 January, within the period covered by our ALMA
observations from 2016 March~8 to 2017 August~18.

The original C- and X-band calibration was performed using the VLA
pipeline in CASA version 4.5.3, while the L- and K-band calibration and
the final imaging of all bands were performed using CASA version
6.7.5.18, following standard VLA continuum calibration procedures.
Standard flux calibrators, 3C~48 or 3C~138, were used for absolute
flux-density calibration. The phase calibrators were J0201$-$1132
for L band, J0215$-$0222 for C and X bands, and J0241$-$0815
for K band.
Both 3C~48 and 3C~138 are standard flux calibrators with known,
slow secular variability. Our CASA reductions accounted for this through
the time-dependent Perley--Butler flux-density models: Perley--Butler
2017 for the L-band data and Perley--Butler 2013 for the C-, X-, and
K-band data. The calibrated flux scale is thus model-based, rather than
a direct measurement of the long-term constancy of either calibrator.
Radio-frequency interference (RFI) was identified and flagged
for each dataset.
The effective central frequency of each image was then recalculated using
the remaining unflagged channels.
All continuum images were produced with CLEAN using natural weighting.

The frequency setups and imaging procedures for the individual VLA
datasets were as follows:
\begin{itemize}

\item
\textit{L band} (project 16A-102, C configuration):
Of the eight SPWs, six (SPW0--SPW5), covering 1.228--2.124~GHz,
were dedicated to continuum observations.
Channels affected by H\,{\sc i} line emission were flagged.
The six SPWs were divided into three sub-bands with bandwidths of
approximately 256~MHz, producing three continuum images per epoch.
To achieve higher sensitivity to the extended emission, we also constructed a deeper
L-band image ($\sigma=0.20$~mJy~beam$^{-1}$) by combining the two observing epochs.

\item
\textit{C band} (project 16B-084, A configuration):
Although the total observed frequency range spanned
3.976--8.023~GHz (SPW2--SPW33), the data above 6.023~GHz
(SPW18--SPW33) were severely degraded by system problems and flagged.
We therefore used the unaffected 3.976--6.023~GHz range
(SPW2--SPW17) to produce a single continuum image.

\item
\textit{X band} (project 16B-084, A configuration):
The continuous frequency range of 7.976--12.024~GHz
($\sim4$~GHz) was divided into four sub-bands with bandwidths of
approximately 1~GHz, producing four continuum images per epoch.  We also combined all SPWs from the two observing epochs to construct a single deeper continuum image ($\sigma=0.008$~mJy~beam$^{-1}$) only for morphological inspection.

\item
\textit{K band} (project 13A-281, C configuration):
The overall frequency range of 17.975--25.767~GHz
($\sim8$~GHz) was divided into two sub-bands with bandwidths of approximately 4~GHz, producing two continuum images.

\end{itemize}

In addition, we used Stokes-$I$ continuum images from the
Very Large Array Sky Survey (VLASS; \citealt{Lacy:2020}) at S band
(2--4~GHz), obtained with the B and BnA configurations.
We median-stacked the three available epochs: 2019 June~21 (VLASS1.2),
2022 January~31 (VLASS2.2), and 2024 August~3 (VLASS3.2).

We measured the nuclear flux density in these images by Gaussian fitting with CARTA \citep{Comrie:2021}.
We also examined morphology to distinguish nuclear from extended emission.  
For the L- and S-band images, in which emission components adjacent to
the nucleus are detected (Section~\ref{section:cm_extended}), we used two-component Gaussian
fitting in CARTA to separate the nucleus from the neighboring
component that could affect the nuclear flux measurement.

Table~\ref{table:vla_observation} summarizes the nuclear flux densities,
effective central frequencies, observation dates, and synthesized beam sizes.
For the 2016 November 27 X-band measurement at 9.51~GHz,
the integrated Gaussian flux density is 1.042~mJy, and its adopted
0.090~mJy uncertainty combines the formal fitting error with a 3\%
flux-calibration term in quadrature.
Previously published pre-upgrade VLA measurements \citep{Doi:2016} are
used only for comparison and are not included in the spectral fitting
described in Section~\ref{section:fit_decomposition}.
This selection was based on the post-upgrade VLA observing era,
rather than on the agreement of individual historical measurements with
the fitted spectrum.

\begin{table*}
\caption{Results of VLA Observations\label{table:vla_observation}}
\begin{center}
\begin{footnotesize}
\begin{tabular}{lcccccc}
\hline
\hline
Date	&	$\nu$	&	$F_\nu$			&	$\sigma$	&	$\theta_\mathrm{maj} \times \theta_\mathrm{min}$			&	$PA$	\\
(yyyy-mm-dd)	&	(GHz)	&	(mJy)			&	(mJy/beam)	&	(arcsec$\times$arcsec)			&	(deg)	\\
(1)	&	(2)	&	(3)			&	(4)	&	(5)			&	(6)	\\
\hline
2013-06-15	&	$20.38$	& $	1.12 \pm 0.12 	$ & 	0.032	& $	1.73	\times	1.08	$ & $	\hphantom{-0}14.0	$ \\
	&	$24.05$	& $	1.06 \pm 0.10 	$ & 	0.025	& $	1.42	\times	0.93	$ & $	\hphantom{-0}14.0	$ \\
2016-03-08	&	$\hphantom{0}1.35$	& $	9.08 	\pm	0.75 	$ & 	1.7	& $	24.54	\times	16.51	$ & $	\hphantom{-0}8.5	$ \\
	&	$\hphantom{0}1.76$	& $	8.74 	\pm	0.55 	$ & 	0.35	& $	19.20	\times	12.77	$ & $	\hphantom{-0}6.5	$ \\
	&	$\hphantom{0}2.00$	& $	7.68 	\pm	0.57 	$ & 	0.25	& $	17.03	\times	11.33	$ & $	\hphantom{-0}6.4	$ \\
2016-04-08	&	$\hphantom{0}1.36$	& $	9.93 	\pm	0.50 	$ & 	0.21	& $	29.71	\times	16.43	$ & $	-22.1	$ \\
	&	$\hphantom{0}1.72$	& $	8.04 	\pm	0.62 	$ & 	0.27	& $	24.56	\times	13.16	$ & $	-23.2	$ \\
	&	$\hphantom{0}1.97$	& $	7.36 	\pm	0.41 	$ & 	0.17	& $	21.23	\times	11.48	$ & $	-21.4	$ \\
2016-11-27	&	$\hphantom{0}8.49$	& $	1.06 	\pm	0.12 	$ & 	0.032	& $	0.42	\times	0.26	$ & $	-34.9	$ \\
	&	$\hphantom{0}9.51$	& $	1.04 	\pm	0.09 	$ & 	0.032	& $	0.38	\times	0.23	$ & $	-34.8	$ \\
	&	$10.48$	& $	0.84 	\pm	0.11 	$ & 	0.029	& $	0.35	\times	0.21	$ & $	-37.2	$ \\
	&	$11.51$	& $	0.84 	\pm	0.12 	$ & 	0.035	& $	0.31	\times	0.19	$ & $	-36.8	$ \\
2016-12-09	&	$\hphantom{0}8.48$	& $	1.09 	\pm	0.09 	$ & 	0.035	& $	0.38	\times	0.23	$ & $	-13.2	$ \\
	&	$\hphantom{0}9.51$	& $	1.13 	\pm	0.06 	$ & 	0.014	& $	0.35	\times	0.21	$ & $	-13.7	$ \\
	&	$10.48$	& $	1.00 	\pm	0.05 	$ & 	0.015	& $	0.32	\times	0.19	$ & $	-15.1	$ \\
	&	$11.51$	& $	0.95 	\pm	0.06 	$ & 	0.019	& $	0.28	\times	0.17	$ & $	-14.2	$ \\
2017-01-15	&	$\hphantom{0}5.07$	& $	1.28 	\pm	0.10 	$ & 	0.035	& $	0.61	\times	0.40	$ & $	-\hphantom{0}8.5	$ \\
VLASS	&	$\hphantom{0}3.00$	& $	3.85 	\pm	0.47 	$ & 	0.090	& $	3.27	\times	2.26	$ & $	-72.9	$ \\
\hline
\end{tabular}
\end{footnotesize}
\end{center}
\begin{flushleft}
\begin{footnotesize}
Col.~(1) observation date;  
Col.~(2) center frequency; 
Col.~(3) flux density; 
Col.~(4) image r.m.s.~noise on blank sky;
Col.~(5)--(6) synthesized beam sizes in major axis, minor axis, and position angle of major axis, respectively.
VLASS flux density is based on a median stack image from three epochs (2019-06-21, 2022-01-31, 2024-08-03).  
\end{footnotesize}
\end{flushleft}
\end{table*}

\section{Results}\label{section:result}
\subsection{The Millimeter-Peaked Spectrum}
\label{section:mm_spectrum}

\begin{figure*}
\begin{center}
\includegraphics[width=0.9\textwidth,trim=10 181 10 218,clip]{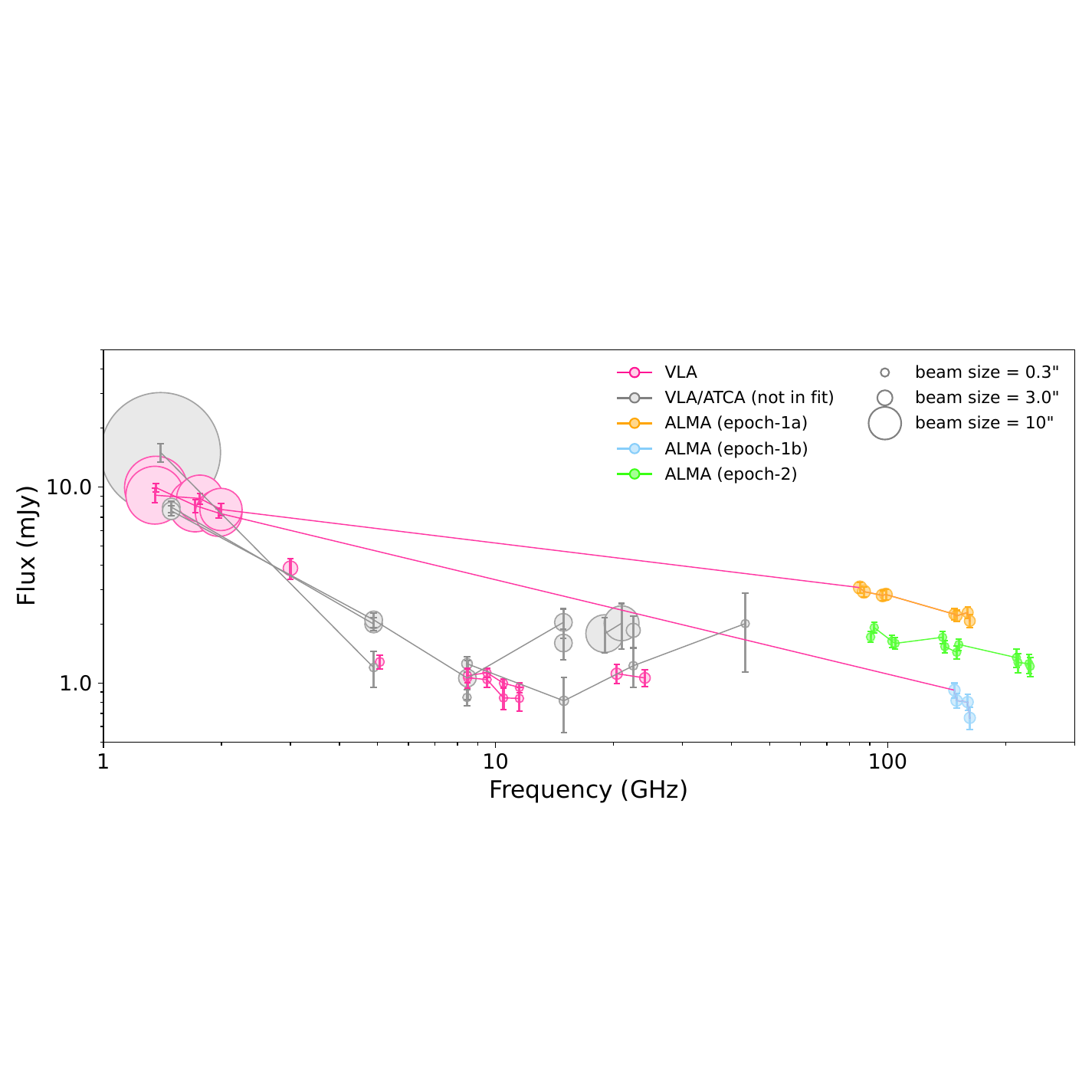}
\end{center}
\caption{
Flux densities of the \target\ nucleus measured with different beam sizes.
Pink symbols show the post-upgrade VLA measurements used in the fit, and
gray symbols show previously published pre-upgrade VLA/ATCA measurements
for comparison only \citep{Doi:2016,Inoue:2018}.
Orange, light-green, and light-blue symbols denote ALMA Epoch-1a, Epoch-2,
and Epoch-1b, respectively.
Lines connect quasi-simultaneous measurements.
Symbol size qualitatively indicates synthesized beam size; representative
beam sizes are shown in the legend.
}
\label{figure:alldataplot}
\end{figure*}

Figure~\ref{figure:alldataplot} presents the NGC~985 nuclear flux densities
versus frequency and the corresponding angular resolutions.  In addition to the archival
post-upgrade VLA measurements presented here for the first time, previously
published pre-upgrade VLA and ATCA measurements are shown for comparison
\citep{Doi:2016}.
These older measurements span several decades; given the
variability of NGC~985, they cannot necessarily be treated as representing
the same spectral state as our ALMA epochs (Section~\ref{section:fit_decomposition}).

Previous quasi-simultaneous multi-frequency observations with the
pre-upgrade VLA revealed an apparent spectral decline between approximately
8 and 15~GHz, followed by a rise toward higher frequencies, suggesting the
presence of a high-frequency excess
\citep{Barvainis:1996,Doi:2016}.
The ALMA Epoch-2 observations subsequently confirmed that this
excess extends into the millimeter regime \citep{Inoue:2018}.

The new ALMA and post-upgrade VLA measurements further support
the presence of a distinct high-frequency component above the extrapolation
of the centimeter-wave spectrum.  The effect of extended emission on the flux densities measured at different angular resolutions is examined in
Section~\ref{section:spatial_separation}, and the spectral decomposition is
presented in Section~\ref{section:fit_decomposition}.

\subsection{Variability of the Millimeter Component}
\label{section:mm_variability}

Figure~\ref{figure:alldataplot} shows flux densities measured with different
beam sizes.  Measured flux densities can be affected by angular-resolution effects in the photometry.  However, the
beam sizes in Epoch-1b are comparable to those in Epoch-1a and substantially
larger than those in Epoch-2, whereas the measured Band~4 flux densities in
Epoch-1b are markedly lower than in the other two epochs.  The decrease from Epoch-1a to Epoch-1b, which are separated by approximately one month, therefore
provides strong evidence for intrinsic variability of a compact millimeter
component.  At comparable Band~4 frequencies, the Epoch-2 flux densities
lie between those of Epoch-1a and Epoch-1b, showing that the millimeter
component was detected again at an intermediate level.

\begin{figure*}
\begin{center}
\includegraphics[width=0.85\textwidth]{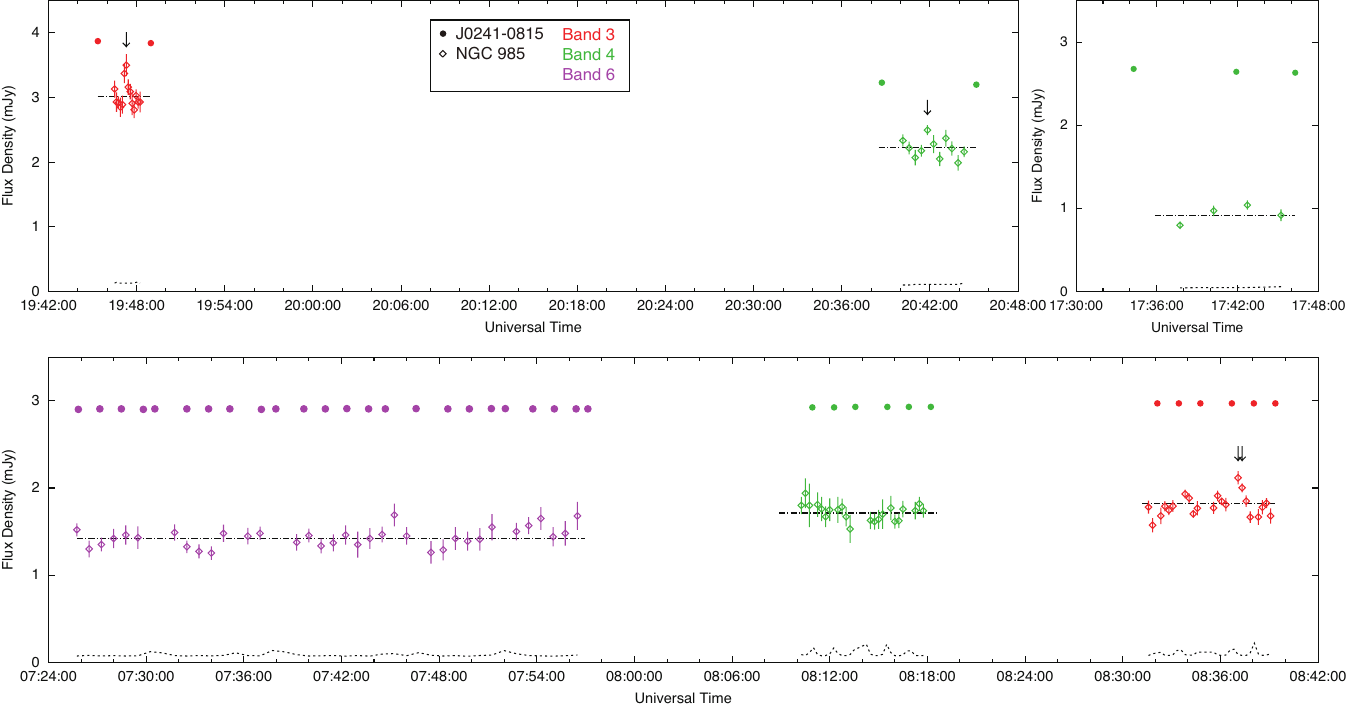}
\end{center}
\caption{
Intrahour ALMA light curves for (a)~Epoch-1a, (b)~Epoch-1b, and
(c)~Epoch-2.
Open diamonds show the \target\ nucleus; filled circles show the phase
calibrator J0241$-$0815, scaled by 0.30\%--0.55\%.
Dashed curves show the image noise, dot-dashed lines the weighted-mean
nuclear flux densities, and arrows mark points deviating by more than $3\sigma$.
}
\label{figure:lightcurve}
\end{figure*}

Figure~\ref{figure:lightcurve} presents the intrahour ALMA light curves.  The flux densities of the phase
calibrator J0241$-$0815, scaled by 0.30\%--0.55\% depending on the band and epoch, are also shown for
comparison.  The measured flux densities of J0241$-$0815 are highly stable over the
intervals between phase-calibrator scans.  Constant-flux fits give
$p=0.037$ and $0.009$ for Band~3 and Band~4 at Epoch-1a,
$p=0.006$ for Band~4 at Epoch-1b, and
$p=3\times10^{-7}$, $0.92$, and $0.43$ for Bands~3, 4, and 6 at Epoch-2,
respectively.  
The fractional rms scatter is 5.4--11.0\% for NGC~985 (median individual \texttt{imfit} errors of 3.7--7.0\%) and 0.03--0.94\% for J0241$-$0815 (median errors of 0.07--0.17\%); the calibrator was sampled only two or three times in Epochs~1a and 1b.
Four individual points deviate by more than $3\sigma$:
19:47:18~UT ($3.0\sigma$) and 20:41:50~UT ($3.3\sigma$) in Epoch-1a,
and 08:37:05~UT ($4.0\sigma$) and 08:37:20~UT ($3.6\sigma$) in Epoch-2.
No individual Epoch-1b point exceeds $3\sigma$.  These excursions occur
on 16--60~s timescales.

These excursions are not secure evidence of intrinsic astrophysical
variability because they could arise from instrumental fluctuations between phase-calibrator scans.  The present
observations therefore do not allow us to establish intrahour variability in
NGC~985.  This should not be interpreted as evidence for its absence; rather, it reflects the limited temporal coverage of the observations.

\subsection{Spatial-Scale Dependence and Extended Emission}
\label{section:spatial_separation}

Since we aim to derive the spectrum of the compact nuclear component of
NGC~985 using observations with different angular resolutions, the
contribution of spatially extended emission must be carefully assessed.
In particular, previous VLA observations have shown that the
measured nuclear flux density depends on angular resolution, suggesting the
presence of spatially extended emission
\citep{Ulvestad:1984,Barvainis:1996,Doi:2016}.

As interferometers, ALMA and the VLA filter out emission on angular scales
larger than the maximum recoverable scale (MRS;
$\sim0.6\lambda/B_{\rm min}$), typically corresponding to
$\sim10$--30 times the synthesized-beam FWHM for commonly used array
configurations.  Structures approaching the MRS may already suffer substantial flux loss \citep[e.g.,][]{Thompson:2017}.  In addition,
low-surface-brightness extended emission may remain undetected because of
limited sensitivity.  
For example, even a $\sim3$~mJy component extended over a region with a diameter of $\gtrsim3$ synthesized-beam FWHMs would have a
characteristic surface brightness of $\lesssim0.3$~mJy~beam$^{-1}$,
corresponding to $\lesssim3\sigma$ in an image with an r.m.s.
noise level of $\sim0.1$~mJy~beam$^{-1}$.  Thus, when the surface brightness of extended emission is comparable to the image noise level, sensitivity limits its detection well before the MRS is reached.

\subsubsection{ALMA Spectra at Different Beam Sizes}
\label{section:ALMA_beamsize}

\begin{figure*}
\begin{center}
\includegraphics[width=\textwidth]{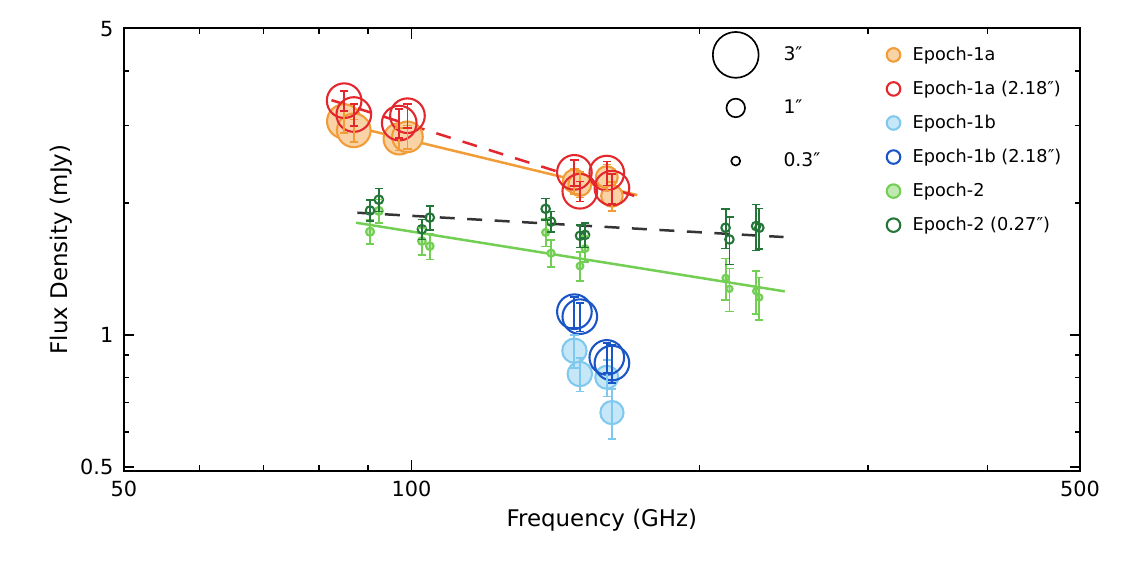}
\end{center}
\caption{
ALMA spectra illustrating beam-size effects.
Filled and open symbols show natural-weighted and beam-matched measurements
from the same visibility datasets; solid and dashed lines are the
corresponding power-law fits. Symbol size indicates synthesized beam
size, scaling linearly from 3~pt at $0.1''$ to 22~pt at $3.0''$.
}
\label{figure:ALMA_beams}
\end{figure*}

Figure~\ref{figure:ALMA_beams} compares the ALMA flux densities measured from the natural-weighted and
beam-matched images produced from the same visibility data.  In Epoch-2, the spectral indices of
the fitted power-law spectra are $-0.12\pm0.06$ and $-0.35\pm0.06$ for the
beam-matched and natural-weighted images, respectively, demonstrating that
the former yields a significantly flatter spectrum.
The spectral-index uncertainties are the formal $1\sigma$ errors from the covariance matrices of the least-squares power-law fits to the ALMA flux densities.
The beam-matched images recover progressively more flux density toward
higher frequencies, which indicates the presence of extended emission that
is not recovered when the beam size varies with frequency.  This rising
trend requires the extended emission to have a flat or inverted spectrum,
rather than the steep spectrum (spectral index $\sim-0.8$) expected for
synchrotron emission associated with star formation.  Possible contributors include free--free
and thermal dust continuum emission.  This possibility is examined further, both in the image analysis below and in the spectral decomposition described in Section~\ref{section:fit_decomposition}.

In Epoch-1a, where the beam sizes are substantially larger than in Epoch-2,
the difference between the measurements from the beam-matched and
natural-weighted images is marginal.  This behavior clearly differs from that in Epoch-2, where the difference is significant, and indicates that the beam sizes in Epoch-1a are already large enough to be sensitive to the extended emission.  The Epoch-1b measurements, obtained
with beam sizes comparable to those of Epoch-1a, do not allow a reliable
evaluation of the spectral index because their frequency coverage is limited
to Band~4.

\subsubsection{Millimeter-Wave Extended Emission}
\label{section:extended_mm}

\begin{figure*}
\begin{center}
\includegraphics[width=0.68\textwidth]{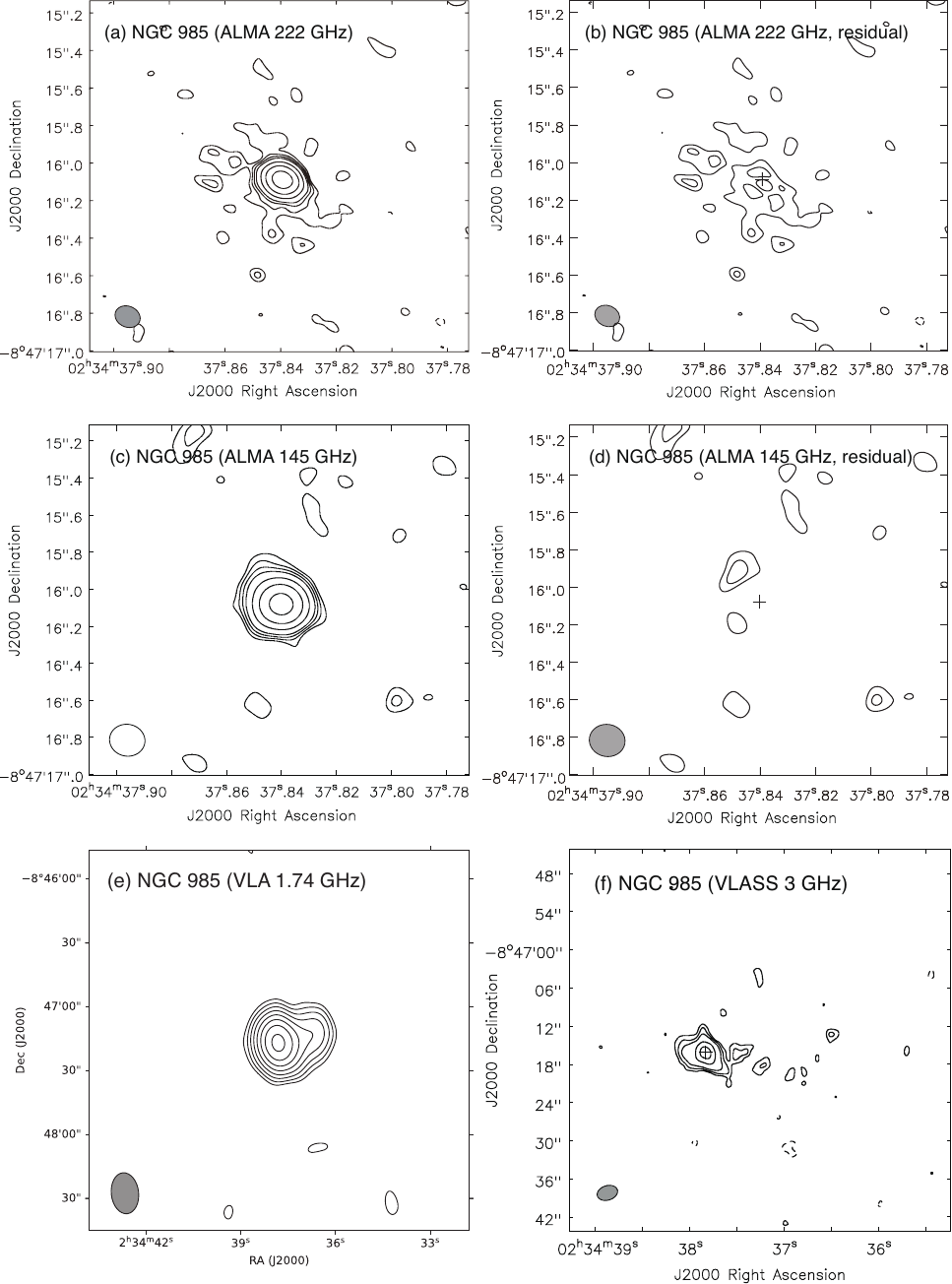}
\end{center}
\caption{
Continuum images and point-source-subtracted residuals of NGC~985.
(a)~ALMA Band~6 at 222~GHz, combining all four SPWs:
contours $3\sigma\times(2/3,\,1,\,\sqrt{2},\,2,\,4,\,8,\,16)$,
$\sigma=0.017$~mJy~beam$^{-1}$, beam
$0.14''\times0.11''$ at $PA=67\fdg0$.
(b)~The same 222-GHz image after subtraction of the fitted point source, revealing faint extended emission; the contour levels and beam are as in (a). The cross marks the subtracted point-source position.
(c)~ALMA Band~4 at 145~GHz, combining all four SPWs:
contours $3\sigma\times(2/3,\,1,\,\sqrt{2},\,2,\,4,\,8,\,16)$,
$\sigma=0.024$~mJy~beam$^{-1}$, beam
$0.19''\times0.17''$ at $PA=79\fdg9$.
(d)~The 145-GHz image after subtraction of the fitted point source, with no significant extended emission remaining around the nucleus; contours are $3\sigma\times(-1,\,2/3,\,1,\,\sqrt{2},\,2,\,4,\,8,\,16)$, with the negative contour dashed. The rms and beam are as in (c). The cross marks the subtracted point-source position.
(e)~VLA L band at 1.74~GHz, combining two epochs:
contours $3\sigma\times(-1,\,1,\,\sqrt{2},\,2,\,4,\,4\sqrt{2},\,8,\,
8\sqrt{2},\,16)$,
$\sigma=0.20$~mJy~beam$^{-1}$, beam
$19.00''\times12.07''$ at $PA=-6\fdg5$.
(f)~Median-stacked VLASS Stokes-$I$ image at 3~GHz from three epochs:
contours $3\sigma\times(-1,\,1,\,\sqrt{2},\,2,\,4,\,8)$,
$\sigma=0.090$~mJy~beam$^{-1}$, beam
$3.27''\times2.26''$ at $PA=-72\fdg9$.
Filled ellipses show the synthesized beams.
}
\label{figure:Images_ALMA_VLAx2}
\end{figure*}

Figure~\ref{figure:Images_ALMA_VLAx2}(a) shows the deep continuum image obtained at the highest ALMA frequency, made by combining all four Band~6 SPWs (representative frequency 222~GHz).  Diffuse emission is
marginally detected at the $\sim2\sigma$ level around the nucleus.
Two-component fitting with a point source and a Gaussian component, using the CASA task \texttt{imfit}, yields flux densities of
$1.21\pm0.02$~mJy for the point source and $0.92\pm0.20$~mJy for the
Gaussian component.  The latter has a deconvolved size of approximately
$0.65''\times0.44''$.
After subtraction of the fitted point source, faint extended emission remains visible (Figure~\ref{figure:Images_ALMA_VLAx2}(b)).

In contrast, only point-like emission was detected in the corresponding
Band~4 image at 145~GHz, constructed by combining all four SPWs
(Figure~\ref{figure:Images_ALMA_VLAx2}(c);
$1\sigma=0.024$~mJy~beam$^{-1}$).
The same two-component fit yielded no significant extended component beyond
the point source, with an upper limit of $0.3$~mJy.
After subtraction of the fitted point source, no significant extended emission remains around the nucleus (Figure~\ref{figure:Images_ALMA_VLAx2}(d)).  The strongly inverted
spectrum implied for the extended component between Bands~4 and 6 suggests that it is attributable to thermal dust emission.

Consequently, a fraction of the extended dust emission could moderately
affect the Band~6 nuclear photometry when a larger beam is used.  Indeed,
as seen in Table~\ref{table:observation}, measurements from the images
restored with a $0.27''$ beam yield flux densities that are higher by approximately 0.46~mJy, on average over the four Band~6 SPWs, than those obtained
with the original synthesized beams of approximately $0.10''$.

\subsubsection{Centimeter-Wave Extended Emission}
\label{section:cm_extended}

In previous radio studies, extended spatial structures in NGC~985 were
clearly resolved by \citet{Appleton:2002} in VLA D-configuration
observations at 4.9 and 8.4~GHz.  They identified two distinct emission
components: a brighter core associated with the Seyfert nucleus and a
secondary component located approximately $17''$ to the west, positioned
just inside the characteristic optical ring structure of the host galaxy.
The western component was interpreted as synchrotron emission presumably
powered by cosmic rays from supernova explosions associated with past
star-forming activity \citep{Appleton:2002}.  This western emission
component could contaminate low-resolution measurements, such as the
1.4-GHz NVSS image obtained with a $\sim45''$ FWHM beam, whereas it should
be spatially separated from the nucleus in VLA~C configuration or more
extended configurations.  On the other hand, no radio counterpart
associated with the companion bulge consisting predominantly of older
stars, located approximately $3\farcs8$ northwest of the Seyfert nucleus
\citep{Perez-Garcia:1996}, was detected either by \citet{Appleton:2002} or
in the present study.

Our L-band image (Figure~\ref{figure:Images_ALMA_VLAx2}(e)) reveals a source
morphology remarkably consistent with that reported by
\citet{Appleton:2002}, clearly detecting both the primary nuclear component
and the western sub-component.  The flux density measurements of the nuclear component were conducted in the present study (Section~\ref{section:VLAdata}).  
Furthermore, in the VLASS S-band image
(Figure~\ref{figure:Images_ALMA_VLAx2}(f)), extended radio emission is marginally
detected at the $\sim2\sigma$ level, spatially coinciding with a
westward-extending tail that forms part of the optical ring structure of
the host galaxy.  The eastern sub-component, with a flux density of
approximately 0.7~mJy and located approximately $3''$ from the nucleus,
is not included in our flux density measurement for the nuclear component.  A hint of this feature
is also visible in the CLEAN residual image of our deep X-band image, although it is not detected at a statistically significant level.

In contrast, the radio emission is identified as a single compact component
centered on the nucleus in all other centimeter-band images, including our
C-, X-, and K-band images in the present study as well as the pre-upgrade VLA
images presented by \citet{Doi:2016}.

\subsection{Spectral Decomposition and Fitting}
\label{section:fit_decomposition}

In our previous study, we decomposed the centimeter--millimeter spectrum
of NGC~985 using both the frequency dependence and the beam-size dependence
of the measured flux densities \citep{Inoue:2018}.  The model consisted of
extended synchrotron and free--free emission associated with the host galaxy,
together with a compact power-law component and synchrotron emission from a
hybrid thermal/nonthermal corona.  We retain this basic framework but modify
it for the present analysis by (i) adopting a Gaussian brightness distribution
for the extended emission, (ii) adding an extended dust component,
(iii) re-evaluating the free--free contribution, (iv) correcting the
hybrid-corona synchrotron calculation, and (v) extending the fitting
procedure to multiple epochs.

For an extended component with a Gaussian brightness distribution, the flux
density enclosed within a beam of size
$\theta_{\rm maj}\times\theta_{\rm min}$ is expressed as
$F_i(\nu)=A_i(\nu/\nu_0)^{\alpha_i}
[1-\exp\{-\theta_{\rm maj}\theta_{\rm min}/\theta_i^2\}]$,
where $\theta_i$ is the effective angular size of component $i$,
$A_i$ is the intrinsic flux density at the reference frequency
$\nu_0=100$~GHz, and $\alpha_i$ is the spectral index.
Here, $\theta_i$ is an effective angular size defined through this
prescription.  \citet{Inoue:2018} instead adopted a uniform-brightness
(top-hat) prescription,
$\min(1,\theta_{\rm maj}\theta_{\rm min}/\theta_i^2)$.
The fitted values of $\theta_i$ are therefore not directly comparable
between the two studies, although the two prescriptions have the same
small- and large-beam limiting behavior.
Following \citet{Inoue:2018}, we fix the spectral index of the synchrotron
emission associated with star-forming activity at
$\alpha_{\rm SF}=-0.8$, characteristic of optically thin synchrotron
emission from star-forming galaxies \citep{Condon:1992}, while its
amplitude $A_{\rm SF}$ and angular size $\theta_{\rm SF}$ are allowed to
vary.

A thermal bremsstrahlung component with $\alpha=-0.1$ was also tested,
following \citet{Inoue:2018}, but did not significantly improve the fit and
introduced additional degeneracy with the other low-frequency components.
We therefore omit an extended free--free component from the final
decomposition.

The extended emission detected at 222~GHz
(Section~\ref{section:extended_mm}) motivates an additional thermal-dust
component.  We adopt a graybody spectrum with a dust emissivity index
$\beta=1.8$ \citep{PlanckCollabolation:2014}, corresponding approximately
to $F_\nu\propto\nu^{3.8}$ in the millimeter regime.  The dust emission is
also represented by a Gaussian brightness distribution, with its amplitude
$A_{\rm dust}$ and effective angular size $\theta_{\rm dust}$ treated as
free parameters.

The compact nuclear emission is represented by a phenomenological
power-law component,
$F_{\rm PL}(\nu)=A_{\rm PL}(\nu/\nu_0)^{\alpha_{\rm PL}}$,
and synchrotron emission from a hybrid thermal/nonthermal corona following
the framework of \citet{Inoue:2014}.  The power-law component is intended primarily to represent unresolved
jet-related radio emission, although other compact continuum contributions,
such as free--free emission, cannot be excluded.  The coronal model is characterized by the magnetic-field
strength $B$, source radius $R$, and nonthermal electron index $p$.
We fix the nonthermal energy fraction at $\eta=0.04$
\citep{Inoue:2008}, and adopt the electron temperature and Thomson optical
depth inferred from the X-ray spectrum, $kT_{\rm e}\simeq29$~keV and
$\tau\simeq3.5$, following the X-ray analysis used previously
\citep{Inoue:2018}.
Here, $\eta$ denotes the fraction of the total coronal electron energy
carried by the nonthermal tail, not the fraction of electrons by number.
The adopted 4\% is a fiducial value motivated by the Seyfert-corona
interpretation of the cosmic MeV gamma-ray background, for which
\citet{Inoue:2008} obtained a nonthermal energy fraction of about 3.5\%,
and its subsequent use in the hybrid-corona synchrotron framework of
\citet{Inoue:2014}. It is not an independent measurement of $\eta$ in
NGC~985. We assess the sensitivity to this assumption in
Section~\ref{sec:disc_alternatives}.  Because the nonthermal power-law tail in this
formulation extends to arbitrarily high electron energies, a finite
nonthermal energy content requires $p>2$.  We therefore impose a lower
bound of $p=2.02$ in the numerical fitting, slightly above the limiting
value $p=2$.

In revisiting the synchrotron calculation, we updated two aspects
of the formulation adopted previously.  First, the nonthermal synchrotron
self-absorption coefficient is evaluated with the standard frequency
dependence $\alpha_\nu\propto\nu^{-(p+4)/2}$ \citep{Rybicki:1979}.
The $\nu^{-(p+5)/2}$ dependence adopted in \citet{Inoue:2014} follows the
expression given in \citet[their equation 21]{Ozel:2000}, which appears to
contain a typographical error in the frequency exponent.  Second, the
normalization of the nonthermal electron contribution now includes the
factor $1/[(p-1)(p-2)]$ consistently in both the emissivity and the
absorption coefficient.  Tests in which these updates were introduced
separately show that the change in the frequency dependence accounts for
most of the resulting shift in the fitted coronal parameters.  The coronal
parameters derived here therefore update those reported in
\citet{Inoue:2018}.

We jointly fit two epoch-specific model spectra for Epoch-1a and Epoch-2.
The compact power-law, star-formation synchrotron, and dust components are
assumed to be common to the two epochs, defining the shared continuum
$F_{\rm shared}=F_{\rm PL}+F_{\rm SF}+F_{\rm dust}$, whereas the coronal
parameters $(p,B,R)$ are fitted independently.  The centimeter-wave
measurements described in Section~\ref{section:VLAdata} do not form
independent epoch-specific datasets corresponding to the two ALMA epochs.
We therefore use the same VLA measurements for both model spectra, dividing
their $\chi^2$ contribution equally between the two branches to avoid
double counting.

Only the measurements from the natural-weighted ALMA images enter the joint
fitting objective; the beam-matched measurements are reserved for the
validation described below.  Two further constraints enter the fit only
once.  The independently measured 222-GHz extended emission,
$0.92\pm0.20$~mJy within the corresponding beam area
(Section~\ref{section:extended_mm}), constrains the dust component through
$\chi^2_{\rm dust}$.  For Epoch-1b, the narrow Band~4 frequency coverage
cannot independently constrain a three-parameter coronal spectrum, so we
do not fit a specific corona.  Instead, the four natural-weighted
measurements are used as one-sided upper constraints on $F_{\rm shared}$,
allowing an arbitrary non-negative compact excess while preventing the
shared continuum from exceeding the observed flux densities.
With correlated calibration uncertainties, our fit objective is formed from
one Gaussian-penalized multiplicative calibration factor $\delta_g$ for
each ALMA observing epoch--receiver band and each VLA observing session.
The same calibration factor applies to all SPWs in a group and is
profiled analytically for a given set of 12 physical parameters. The
independently measured 222-GHz extended-emission constraint shares the
Epoch-2 Band-6 calibration factor with the natural-weighted Band-6 fluxes.
The six VLA observing-session factors are shared between the two model branches; each group incurs
only one Gaussian calibration penalty. The VLASS stacked flux density,
which spans three observing epochs, retains its tabulated error rather than
a single-session common calibration factor. The one-sided Epoch-1b
upper constraints are profiled with a non-negative excess and a common
Band-4 calibration factor. In schematic notation the penalized objective is
$Q=\chi^2_{\rm VLA,1a}/2+\chi^2_{\rm ALMA,1a}+
\chi^2_{\rm VLA,2}/2+\chi^2_{\rm ALMA,2}+
\chi^2_{\rm dust}+\chi^2_{\rm 1b,UL}+
\sum_g (\delta_g/\sigma_{{\rm cal},g})^2$,
where the observational residuals use the formal statistical errors after
profiling the shared calibration factors. The calibration-fraction uncertainties are 5\%
for ALMA Bands~3/4, 10\% for ALMA Band~6, 5\% for the VLA K band, and 3\%
for the remaining individual VLA observing sessions. The objective uses
43 distinct observational constraints and 12 physical fit parameters,
in addition to the profiled calibration factors. Repeated least-squares
minimizations from different initial conditions give a minimum
$Q_{\rm min}=27.27$. This penalized, half-weighted objective is not a
conventional chi-square statistic with uniquely defined degrees of
freedom, and we do not quote a reduced $\chi^2$.

To quantify nonlinearity and parameter covariance we perform 1000
parametric Monte Carlo trials. Each realization includes independent
Gaussian noise drawn from the formal Gaussian-fit flux-density errors
and one common multiplicative calibration perturbation for every ALMA
epoch--band or VLA observing session. The extended 222-GHz flux constraint
shares its calibration perturbation with ALMA Epoch-2 Band~6; VLA fluxes
are perturbed once and reused in both half-weight branches. The same
profiled-calibration objective is reoptimized for all 12 physical
parameters in every trial. Table~\ref{table:spectral_fit} reports the
least-squares best-fit parameter values and the offsets to the
16th--84th percentile boundaries of the Monte Carlo best-fit distributions.
These uncertainties refer to the adopted model and calibration treatment,
and do not include uncertainty associated with the fixed nonthermal
energy fraction $\eta$.

\begin{table}
\caption{Spectral Fitting Results for NGC 985}
\label{table:spectral_fit}
\begin{center}
\begin{tabular}{lcc}
\hline
Parameter & Epoch-1a & Epoch-2 \\
(1) & (2) & (3) \\
\hline
\multicolumn{3}{l}\textit{Hybrid corona} \\
$p$ & $3.16^{+0.53}_{-0.43}$ & $2.21^{+0.62}_{-0.18}$ \\
$B$ (G) & $2.67^{+2.68}_{-1.20}$ & $0.814^{+0.771}_{-0.015}$ \\
$R$ ($R_{\rm S}$) & $327^{+80}_{-54}$ & $192^{+65}_{-8}$ \\
$\eta$ & $0.04$ (fixed) & $0.04$ (fixed) \\
\hline
\multicolumn{3}{l}\textit{Shared compact power-law component} \\
$A_{\rm PL}$ (mJy) & \multicolumn{2}{c}{$0.366^{+0.066}_{-0.092}$} \\
$\alpha_{\rm PL}$ & \multicolumn{2}{c}{$-0.417^{+0.069}_{-0.121}$} \\
\hline
\multicolumn{3}{l}\textit{Shared star-formation synchrotron component} \\
$A_{\rm SF}$ (mJy) & \multicolumn{2}{c}{$0.245^{+0.012}_{-0.019}$} \\
$\alpha_{\rm SF}$ & \multicolumn{2}{c}{$-0.8$ (fixed)} \\
$\theta_{\rm SF}$ (arcsec) & \multicolumn{2}{c}{$3.00^{+0.61}_{-0.39}$} \\
\hline
\multicolumn{3}{l}\textit{Shared dust component} \\
$A_{\rm dust}$ (mJy) & \multicolumn{2}{c}{$0.0542^{+0.0115}_{-0.0139}$} \\
$\theta_{\rm dust}$ (arcsec) & \multicolumn{2}{c}{$0.420^{+0.114}_{-0.238}$} \\
\hline
\end{tabular}
\end{center}
\vspace{1mm}
\begin{minipage}{0.94\columnwidth}
\footnotesize
Col.~(1) fitted parameter; Col.~(2) Epoch-1a result; Col.~(3) Epoch-2 result.
Uncertainties indicate offsets from the best-fit values to the 16th--84th
percentile boundaries of 1000 parametric Monte Carlo realizations (Section~\ref{section:fit_decomposition}).
The amplitudes $A_{\rm PL}$, $A_{\rm SF}$, and
$A_{\rm dust}$ are defined at $\nu_0=100$~GHz.
For Epoch-2, 11.8\% of the realizations reach the imposed lower bound $p=2.02$.
\end{minipage}
\end{table}

The resulting best-fitting coronal parameters are
$(p,B,R)=(3.16,2.67~{\rm G},327R_{\rm S})$ for Epoch-1a and
$(2.21,0.814~{\rm G},192R_{\rm S})$ for Epoch-2. Of the 1000
Monte Carlo fits, 92.1\% have $p_{\rm 1a}>p_2$, 94.4\% have
$B_{\rm 1a}>B_2$, and 94.2\% have $R_{\rm 1a}>R_2$.
These fractions are conditional on the common centimeter-wave constraints
and the adopted spectral model, and should not be interpreted as
independent significance tests of physical evolution.
The Epoch-2 electron index reaches the imposed lower bound $p=2.02$
in 11.8\% of the realizations, yielding a strongly asymmetric uncertainty.
The turnover remains poorly sampled, and $p$, $B$, and $R$ are
strongly covariant, as discussed in Section~\ref{sec:disc_corona}.

When all 17 historical measurements (Section~\ref{section:mm_spectrum}) are included in the joint fit, the contribution of the adopted measurements to the fitting objective increases from 27.27 to 52.28, and the Epoch-2 magnetic field becomes poorly constrained. This deterioration may reflect a combination of intrinsic variability in the centimeter-wave emission and heterogeneous observational characteristics.

A diagnostic Markov chain Monte Carlo (MCMC) calculation with the same joint
fitting setup visualizes the covariance among $p$, $B$, and $R$; the corner
plots appear in
Appendix~\ref{section:mcmc_appendix}.  Parameter uncertainties in
Table~\ref{table:spectral_fit} are instead taken from the parametric Monte
Carlo analysis described above.

\begin{figure*}
\begin{center}
\includegraphics[width=0.9\textwidth,trim=10 138 10 175,clip]{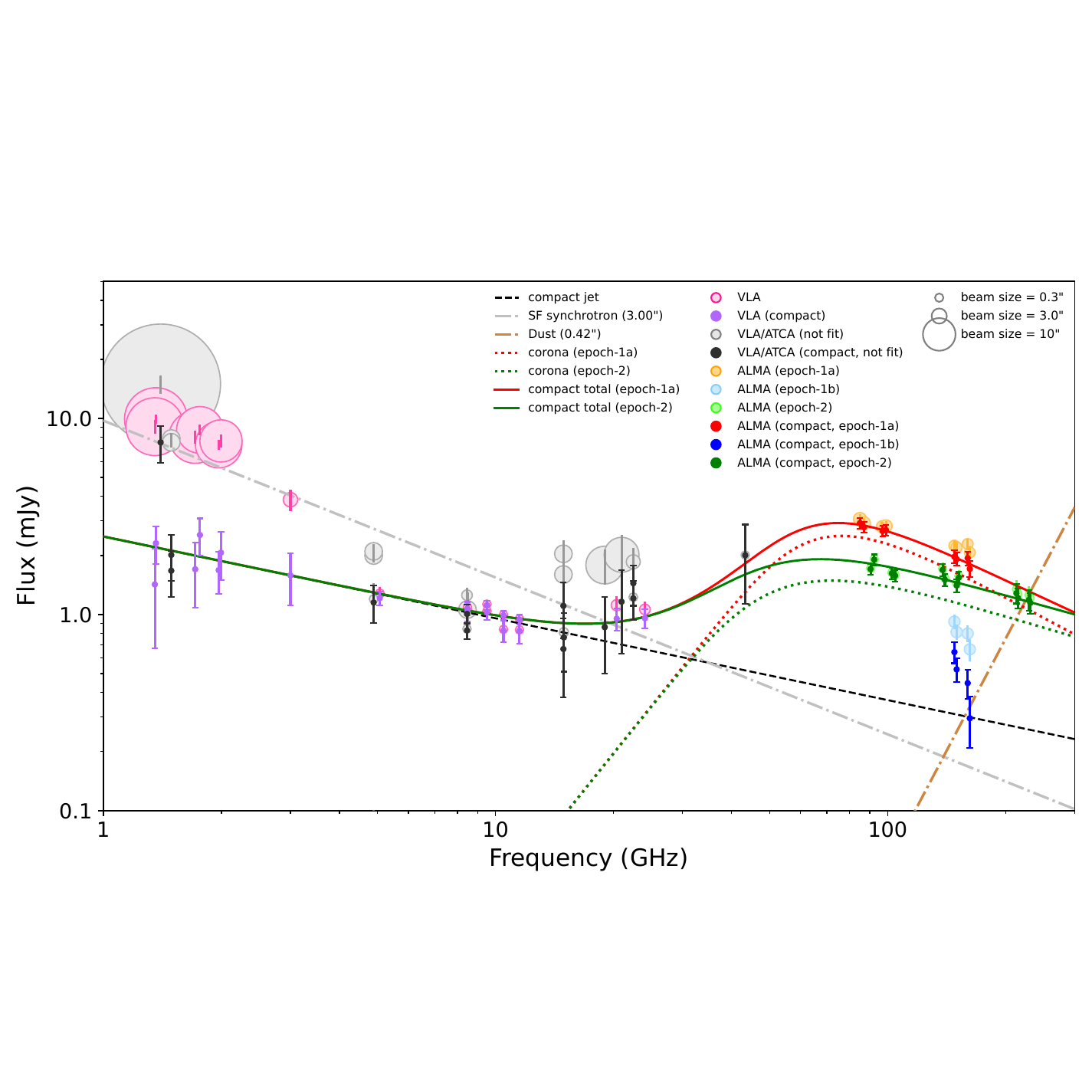}
\end{center}
\caption{
Centimeter--millimeter spectrum of \target\ showing the compact nuclear
component after subtraction of the fitted extended synchrotron and dust
emission.
For the post-upgrade data, lighter and darker symbols show raw and
subtracted flux densities, respectively: VLA (pink/purple), Epoch-1a
(orange/red), Epoch-1b (light-blue/blue), and Epoch-2 (light-green/green).
Gray/black symbols show pre-upgrade VLA/ATCA data for comparison only.
The dashed line is the shared compact power law; gray and brown dot-dashed
lines are the extended synchrotron and dust components.
Dotted and solid red/green lines show the hybrid-corona component and total
compact spectrum for Epoch-1a/Epoch-2.
Epoch-1b provides one-sided upper constraints only.
}
\label{figure:spectraldecomposition}
\end{figure*}

Using the best-fit solution, we subtract the fitted extended synchrotron and
dust contributions to obtain the compact nuclear spectra shown in
Figure~\ref{figure:spectraldecomposition}.  The hybrid-corona model
reproduces the Epoch-1a and Epoch-2 radio--millimeter spectra well.
The relatively high residual flux density of the NVSS point near 1.4~GHz may reflect more extended emission that is not captured by the fitted Gaussian synchrotron component with $\theta_{\rm SF}\simeq3.00''$.

\begin{figure*}
\begin{center}
\includegraphics[width=0.85\textwidth]{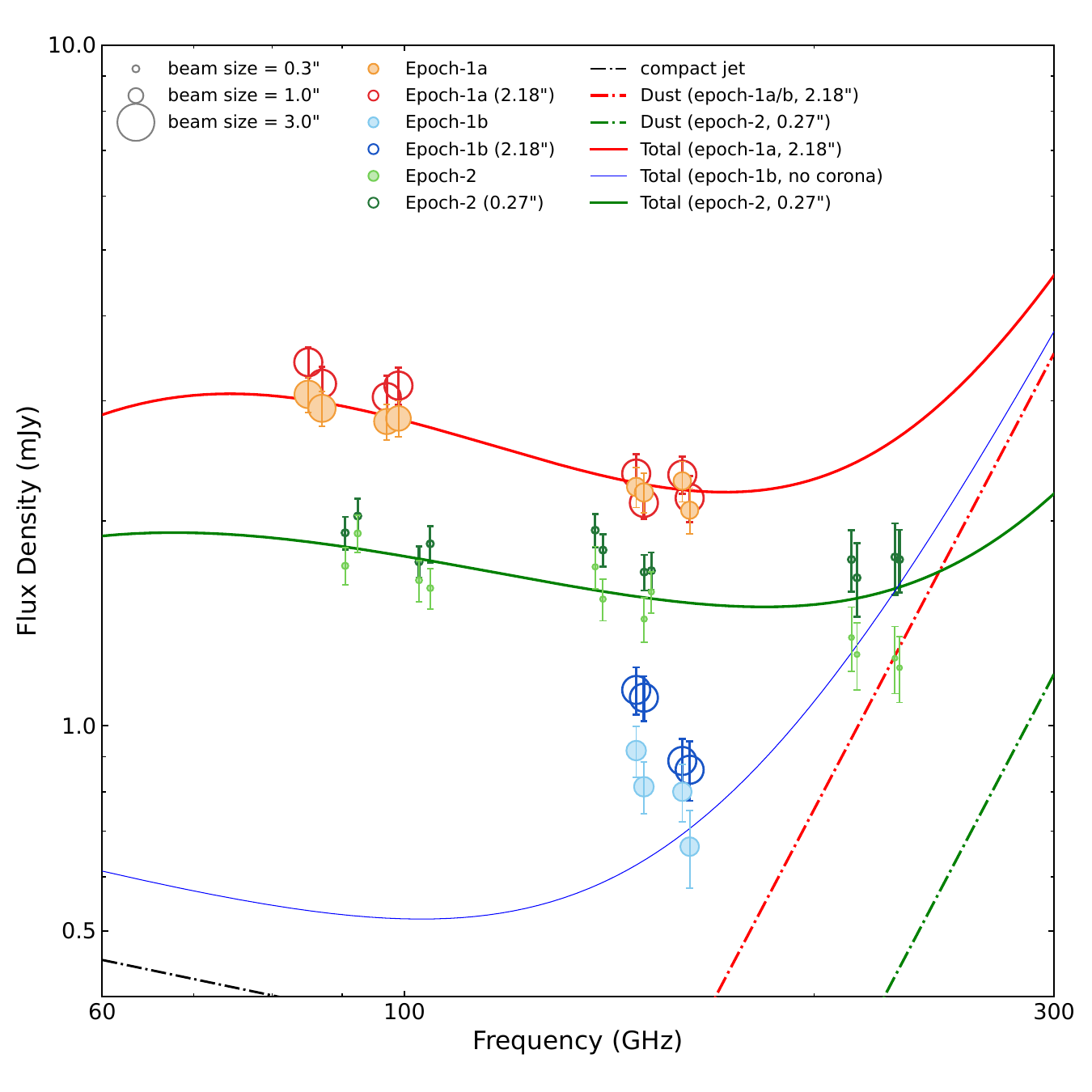}
\end{center}
\caption{
ALMA spectra illustrating the spectral decomposition.
 Epoch-1a (orange/red), Epoch-1b (light-blue/blue), and Epoch-2
(light-green/green). Filled and open symbols show natural-weighted
and beam-matched measurements, with the same symbol-size scaling as in
Figure~\ref{figure:ALMA_beams}. The dot-dashed black line is the shared
compact power law, colored dot-dashed lines the fitted dust contribution,
solid red/green lines the total model for Epoch-1a/Epoch-2, and the thin
blue line the shared Epoch-1b continuum. Beam-matched measurements are not
included in the fit.
}
\label{figure:ALMA_decomposition}
\end{figure*}

Figure~\ref{figure:ALMA_decomposition} provides an additional validation of the
spatial decomposition using the beam-matched ALMA measurements.  In
Epoch-1a, the fitted dust contribution is modest over Bands~3 and 4,
consistent with the small beam-size dependence of the measured flux
densities.  In Epoch-2, the same dust model becomes important toward
Band~6 and reproduces the enhanced flux densities measured with the larger
$0.27''$ beam.  Thus, the fitted dust component also accounts
for a beam-size dependence that was not used directly to determine the
best-fitting parameters.

\section{Discussion}
\label{sec:discussion}

The ALMA observations reveal a recurrent and variable compact millimeter
component in NGC~985.  It is independently detected in Epoch-1a, becomes
much weaker approximately one month later in Epoch-1b, and is detected again
at an intermediate level in Epoch-2.  We discuss below the constraints from
this variability and its implications for the origin of the compact
millimeter emission.

\subsection{A Recurrent and Variable Compact Millimeter Component}
\label{sec:disc_variability}

NGC~985 has long been known to exhibit a high-frequency excess above the extrapolation of its steep-spectrum centimeter-wave emission \citep{Antonucci:1988,Barvainis:1996,Doi:2016}. The ALMA Epoch-2 observation established that this excess extends into the millimeter regime and can be separated from the more extended host-galaxy emission \citep{Inoue:2018}. The present Epoch-1a observation independently recovers the compact millimeter excess, whereas the much lower Epoch-1b flux density shows that its strength can change substantially on a timescale of approximately one month. The intermediate state observed in Epoch-2 further indicates that the component is recurrent or state-changing rather than strictly persistent.

The month-scale change provides the secure variability signal used in the
following discussion.  The intrahour light curves contain several isolated
excursions (Section~\ref{section:mm_variability}), but these are not
sufficiently robust to establish intrinsic short-timescale variability.
Such short-timescale searches remain physically motivated
because the radiative cooling times of the
millimeter-emitting electrons can be much shorter than the sampling
interval of the present observations (Section~\ref{sec:disc_corona}).

\subsection{Compactness Implied by the Month-Scale Variability}
\label{sec:disc_compactness}

The approximately 30-day interval between Epoch-1a and Epoch-1b provides
a causality upper limit on the size of the variable emitting region.
Correcting to the rest frame gives
$R_{\rm var}\lesssim c\,\Delta t_{\rm obs}/(1+z)\simeq0.024$~pc, or
$\simeq1.1\times10^3R_{\rm S}$,
adopting $M_{\rm BH}=2.2\times10^8M_\odot$, the same
value used by \citet{Doi:2016} with reference to
\citet{Kim:2008} and \citet{Vasudevan:2009}.
Because the two observations only bracket the transition, the actual
variability timescale, and hence the source size, may be substantially
smaller.

This constraint independently requires the variable millimeter emission to
arise from a compact nuclear region rather than from the kiloparsec-scale
galactic synchrotron or extended dust components identified in the imaging
and spectral decomposition.  The size limit is also consistent with the
characteristic radii obtained from the hybrid-corona fits in
Table~\ref{table:spectral_fit}, although it was not imposed as a prior.
Denser monitoring of other radio-quiet AGNs has yielded tighter causality
limits from day-scale millimeter variability
\citep{Michiyama:2024,Shablovinskaya:2024}; the less restrictive constraint
for NGC~985 mainly reflects the wider temporal sampling of the present
observations.

\subsection{Spectral Changes in the Hybrid-Corona Interpretation}
\label{sec:disc_corona}

Within the hybrid-corona model, the two well-sampled ALMA epochs favor
different spectral states.  The best-fitting nonthermal electron indices are
$p_{\rm 1a}=3.16$ and $p_2=2.21$, with
$p_{\rm 1a}>p_2$ in 92.1\% of the Monte Carlo realizations.  For optically
thin synchrotron emission, the spectral slope is related to the electron
index through $\alpha\simeq-(p-1)/2$, and in the present fits $p$ is
constrained mainly by the high-frequency side of the millimeter component.
The higher-flux Epoch-1a state therefore favors a steeper nonthermal electron
distribution than Epoch-2.

The fitted parameters also allow order-of-magnitude estimates of the
timescales relevant to these spectral states. For
$M_{\rm BH}=2.2\times10^8M_\odot$, the fitted radii correspond to
light-crossing times $R/c\approx8$ and 5~days for Epoch-1a and Epoch-2,
respectively. Electrons radiating at 100~GHz have Lorentz factors
$\gamma_e\approx[\nu/(4.2\times10^6B~{\rm Hz})]^{1/2}\approx90$ and 170 in
the two epochs, for which the synchrotron cooling times are approximately
13 and 80~days. Inverse-Compton cooling can be considerably faster.
Approximating the radiation energy density as
$U_{\rm rad}\approx L_{\rm eff}/(4\pi R^2c)$ and adopting illustrative
effective luminosities $L_{\rm eff}=10^{44}$--$10^{45}$~erg~s$^{-1}$ for
optical--UV seed photons from the accretion disk, the combined cooling times
are approximately 15~hr--4~days for Epoch-1a and 3~hr--1~day for Epoch-2.
The upper value corresponds to the bolometric luminosity of NGC~985
estimated with typical hard-X-ray bolometric corrections
\citep{Vasudevan:2009}, and the lower value allows for a reduced effective
photon density at the fitted radius owing to geometry and anisotropy. These estimates
depend on the geometry and spectrum of the irradiating photon field and,
through the $\eta$--$B$--$R$ degeneracy (Section~\ref{sec:disc_alternatives}),
on the adopted $\eta$; they are not measurements of the electron cooling
times. They nevertheless indicate that the cooling times are likely shorter
than, or comparable to, the light-crossing time, so that the millimeter
emission must be sustained by continuous injection or reacceleration of
nonthermal electrons, as in coronal models in which particle acceleration
balances radiative cooling \citep{Inoue:2019}.

The change in the electron index between Epoch-1a and Epoch-2 could reflect differences in the injected nonthermal
population, the balance between particle acceleration and radiative cooling
\citep[e.g.,][]{Inoue:2019,Ly:2026}, or the spatial mixture
of emitting zones \citep[e.g.,][]{Raginski:2016}.  Given the timescales
above, the local electron distribution responds to a change in injection or
acceleration within hours to days, whereas the integrated flux density
responds on the light-crossing time or on longer timescales for changes
propagating across the region; both are compatible with the observed
month-scale change. The difference in $p$, however, is not explained simply
by stronger radiative cooling in the higher-flux state: for a common photon
field, the electrons radiating at ALMA frequencies cool faster relative to
$R/c$ in Epoch-2, which nevertheless shows the flatter distribution. The Epoch-2 fit lies close to the
imposed lower bound $p=2.02$, and the 95\% Monte Carlo interval of
$\Delta p=p_{\rm 1a}-p_2$ includes zero.  We therefore regard the difference
in $p$ as a tendency in the spectral shape rather than a formal
high-significance measurement of a change in the electron distribution.

The fitted characteristic radius is also larger in Epoch-1a than in Epoch-2, with best-fitting values of $327R_{\rm S}$ and $192R_{\rm S}$, respectively.
  The synchrotron turnover between the VLA K band and the ALMA
bands is not directly sampled, however, so the turnover frequency, $R$, and
$B$ remain strongly coupled.  In the one-zone model, part of the higher
Epoch-1a millimeter flux is accommodated by a larger effective emitting area;
we therefore regard $R$ primarily as an effective size parameter rather than
as direct evidence for physical expansion of the corona.  The fitted magnetic field is similarly coupled to the turnover and to $p$, and its epoch-to-epoch difference is not interpreted separately.

Overall, the multi-epoch spectra favor a change in the state of the compact
synchrotron-emitting plasma, including its nonthermal electron distribution,
rather than a simple multiplicative change in flux density.

\subsection{Parameter Covariance and Alternative Compact Synchrotron Origins} \label{sec:disc_alternatives} 

The physical parameters of a one-zone self-absorbed synchrotron model are strongly coupled when the turnover is not well sampled. The diagnostic MCMC analysis (Appendix~\ref{section:mcmc_appendix}) illustrates this directly: $p$, $B$, and $R$ occupy elongated and curved distributions rather than varying independently. The fitted values of $B$ and $R$ should therefore be interpreted together with the other model parameters and the assumed homogeneous geometry.

To assess the additional degeneracy associated with the assumed
nonthermal energy fraction, we repeated the full joint fit with a common
$\eta=0.01$, 0.04, and 0.10 in both epochs, reoptimizing the other 12
parameters and the common calibration nuisance factors in each case.
The profiled objective remains $\chi^2_{\rm obj}\simeq27.27$ and the
model spectra are practically indistinguishable. The fitted electron
indices stay near $p_{\rm 1a}=3.163$ and $p_2=2.208$. In contrast,
as $\eta$ increases from 0.01 to 0.10, the fitted magnetic fields change
from 4.50 to 1.86~G (Epoch-1a) and from 1.59 to 0.513~G (Epoch-2),
while the effective radii change from $372R_{\rm S}$ to $299R_{\rm S}$
and from $227R_{\rm S}$ to $171R_{\rm S}$, respectively. This near-invariance of the spectra and
$p$, accompanied by substantial changes in $B$ and $R$, illustrates the
$\eta$--$B$--$R$ degeneracy \citep[see also][]{Inoue:2019}. Therefore,
the quoted $B$ and $R$ values are conditional on the assumed
$\eta=0.04$. This sensitivity test does not constrain possible
variations in $\eta$ between epochs or modify the statistical
uncertainty of the epoch-to-epoch difference in $p$.

As a phenomenological check, we replaced the hybrid-corona component with a homogeneous synchrotron self-absorption spectrum of a single power-law electron population without a thermal component, having the same number of free parameters. The resulting fits are nearly indistinguishable at the present frequency sampling. The radio--millimeter SED therefore requires a compact curved, self-absorbed synchrotron component, but does not uniquely identify its physical origin with the compact X-ray-producing corona. 

A radially stratified corona, in which the magnetic field and particle
properties vary with radius, can produce a flatter synchrotron spectrum
than a homogeneous one-zone source \citep{Raginski:2016}, while a magnetically connected
coronal outflow can place much of the millimeter-emitting plasma outside
the most compact X-ray region while retaining a connection to the
accretion flow \citep{Hankla:2026}.
A weak compact jet or jet-base component \citep[e.g.,][]{Markoff:2005} provides another possible origin of the compact synchrotron emission.  In these inhomogeneous scenarios, the observed millimeter spectral curvature could arise from the superposition of synchrotron-emitting regions with different physical conditions, rather than from a single homogeneous corona. Consequently, the parameters inferred from the one-zone hybrid-corona fits would not necessarily represent the physical properties of an individual emitting region. The present results are therefore consistent with a magnetized-corona interpretation, while allowing the millimeter-emitting region to be spatially distinct from the X-ray Comptonizing plasma.

\subsection{Comparison with the Hard-X-Ray Variability}
\label{sec:disc_bat}

We also examined the monthly Crab-weighted Swift/BAT hard-X-ray light curve
from the 157-month survey \citep{Lien:2025} around the three ALMA epochs,
using the monthly binning as a compromise between sensitivity and temporal
resolution.  No clear correspondence is apparent between the millimeter
states and the BAT flux.  The individual monthly BAT measurements have
limited significance and average over timescales different from the ALMA
observations, so the present data do not support a clear positive or inverse
millimeter--X-ray correlation.

Population studies have found a relatively tight relation between compact nuclear millimeter and X-ray luminosities in radio-quiet AGNs \citep{Kawamuro:2022,Ricci:2023}, indicating a close connection between millimeter emission and accretion power.
We estimate the compact 100-GHz flux densities from the adopted
shared power-law component plus the epoch-specific coronal components,
obtaining 2.653 and 1.754~mJy for Epochs~1a and 2, respectively.
For Epoch-1b, the shared compact power law alone gives 0.366~mJy, an
illustrative value that assumes no contribution from a separately
unconstrained coronal component. Using the absorption-corrected
14--150~keV flux of $2.68\times10^{-11}$~erg~cm$^{-2}$~s$^{-1}$ from the
BASS 70-month spectral analysis \citep{Ricci:2017}, the resulting values of
$\log_{10}[\nu L_\nu(100\,\mathrm{GHz})/L_{14-150\,\mathrm{keV}}]$
are $-4.01$, $-4.88$, and $-4.20$ for Epochs~1a, 1b, and 2,
respectively. At the adopted distance, these are $+0.69$, $-0.17$,
and $+0.51$~dex relative to the luminosity regression of
\citet{Ricci:2023}. Because the X-ray luminosity is a non-simultaneous
long-term average, these values provide reference comparisons rather
than contemporaneous millimeter--X-ray ratios.

Short-timescale monitoring of individual sources, however, has yielded mixed or inconclusive correlations \citep{Behar:2020,Petrucci:2023,Shablovinskaya:2024}. In particular, IC~4329A showed a factor-of-three millimeter variation within several days without a significant correlation with the X-rays on timescales below 20~days \citep{Shablovinskaya:2024}. The behavior of NGC~985 is therefore consistent with a picture in which the population-level millimeter--X-ray connection does not necessarily imply one-to-one tracking on short timescales.

\subsection{Summary and Physical Implications}
\label{sec:disc_context}

Millimeter variability is now observed in a growing number of radio-quiet
AGNs, including NGC~7469, MCG+08-11-11, GRS~1734$-$292, IC~4329A,
and NGC~1566
\citep{Baldi:2015,Behar:2020,Petrucci:2023,Michiyama:2024,
Shablovinskaya:2024,Jana:2025,Venselaar:submitted}.  NGC~985 adds a source with a long-established
high-frequency excess in which the compact millimeter component is detected
in multiple ALMA epochs and changes substantially in strength.  The
multi-frequency coverage further allows the spectral shapes of different
millimeter states to be compared within a common decomposition framework.

Our results can be summarized at three levels of model dependence.
(i) Independently of the spectral model, a compact millimeter excess is
detected in all three ALMA epochs; it decreased substantially within
approximately one month and was detected again at an intermediate level,
which limits the variable emitting region to
$\lesssim1.1\times10^3R_{\rm S}$ (Section~\ref{sec:disc_compactness}).
(ii) The spectral decomposition requires a compact, curved, self-absorbed
synchrotron component in addition to extended star-formation synchrotron
and dust emission, although a homogeneous self-absorbed synchrotron
spectrum from purely nonthermal electrons describes it equally well
(Section~\ref{sec:disc_alternatives}). (iii) Within the one-zone
hybrid-corona model, the two well-sampled epochs correspond to different
spectral states: the higher-flux Epoch-1a state favors a steeper nonthermal
electron distribution, and the fitted magnetic field and effective radius
also differ (Table~\ref{table:spectral_fit}). These differences in $B$ and $R$ are model-dependent (Sections~\ref{sec:disc_corona} and \ref{sec:disc_alternatives}) and do not by themselves demonstrate changes in field strength or size, although genuine changes are
not excluded.

Because the millimeter-emitting electrons must be continuously supplied (Section~\ref{sec:disc_corona}), we consider changes in the injection
rate or injected spectrum within a compact magnetized region to be the most
natural driver of the observed changes in both the strength and the
high-frequency slope, without requiring a global change in field strength or
size. Alternatively, in a stratified corona, a corona-connected outflow, or a
jet base, similar changes can arise from the varying contributions of zones
with different physical conditions (Section~\ref{sec:disc_alternatives}); the
present data do not distinguish between these possibilities.

Taken together, a magnetized hybrid corona
provides a plausible physical origin of the compact millimeter
component in NGC~985, while a stratified corona,
corona-connected outflow, or weak compact jet remains compatible with the
present radio--millimeter spectra.  
Broader simultaneous frequency coverage across the VLA--ALMA gap would
directly resolve the synchrotron turnover and test whether the low-frequency spectral shape is
consistent with a homogeneous one-zone source or instead reflects radial
structure in the emitting plasma, and denser millimeter and X-ray monitoring
would determine the variability timescale and test how closely the compact
millimeter emission tracks the high-energy source.

\begin{acknowledgments}
This work was supported in part by JSPS KAKENHI Grant Number
JP20K04020 (A.D.).
Y.I. is supported by JST FOREST Program (Grant Number
JPMJFR256L, Japan), MEXT SPReAD Program (Grant Number
JPMXP1726291302, Japan), and JSPS KAKENHI Grant Numbers JP22K18277,
JP26H00604, and JP26K22331.

This paper makes use of the following ALMA data:
ADS/JAO.ALMA\#2015.1.01516.S and
ADS/JAO.ALMA\#2016.1.01140.S.
ALMA is a partnership of ESO (representing its member states), NSF (USA)
and NINS (Japan), together with NRC (Canada), NSTC and ASIAA (Taiwan),
and KASI (Republic of Korea), in cooperation with the Republic of Chile.
The Joint ALMA Observatory is operated by ESO, AUI/NRAO and NAOJ.

The National Radio Astronomy Observatory and Green Bank Observatory are
facilities of the U.S. National Science Foundation operated under cooperative
agreement by Associated Universities, Inc.

We acknowledge the use of public data from the Swift data archive.

During the preparation of this work, the authors used ChatGPT (OpenAI)
for assistance with manuscript editing and the development and revision
of analysis code.  The authors reviewed and verified all AI-assisted
output and take full responsibility for the content of this work.
\end{acknowledgments}

\facilities{ALMA, VLA, Swift}
\software{CASA \citep{McMullin:2007}, CARTA \citep{Comrie:2021}}

\appendix

\section{Diagnostic MCMC Analysis}
\label{section:mcmc_appendix}

We ran a diagnostic 48-walker ensemble MCMC for 10,000 steps, discarding
the first 3,000 steps as burn-in, to visualize the covariance of the
hybrid-corona parameters within the current profiled-calibration joint
objective of Section~\ref{section:fit_decomposition}. The chain targets
$\exp(-Q/2)$ with broad uniform priors in physical parameters. At each
step the correlated calibration nuisance factors are analytically profiled,
not marginalized: this is therefore a diagnostic pseudo-posterior, not a
fully calibrated Bayesian posterior. The mean acceptance fraction is
0.126; the slowest inferred autocorrelation time is approximately 646
steps and the corresponding retained-chain-length ratio is only
$N/\tau\simeq10.8$, below the usual standard for reliable posterior
uncertainties. Accordingly, MCMC is used only to illustrate parameter
covariance; Table~\ref{table:spectral_fit} uncertainties are based on the
1000-trial correlated-calibration parametric Monte Carlo described above.

\begin{figure}[!htb]
\begin{center}
  \begin{minipage}[t]{0.49\textwidth}
    \centering
    \includegraphics[width=\linewidth]
    {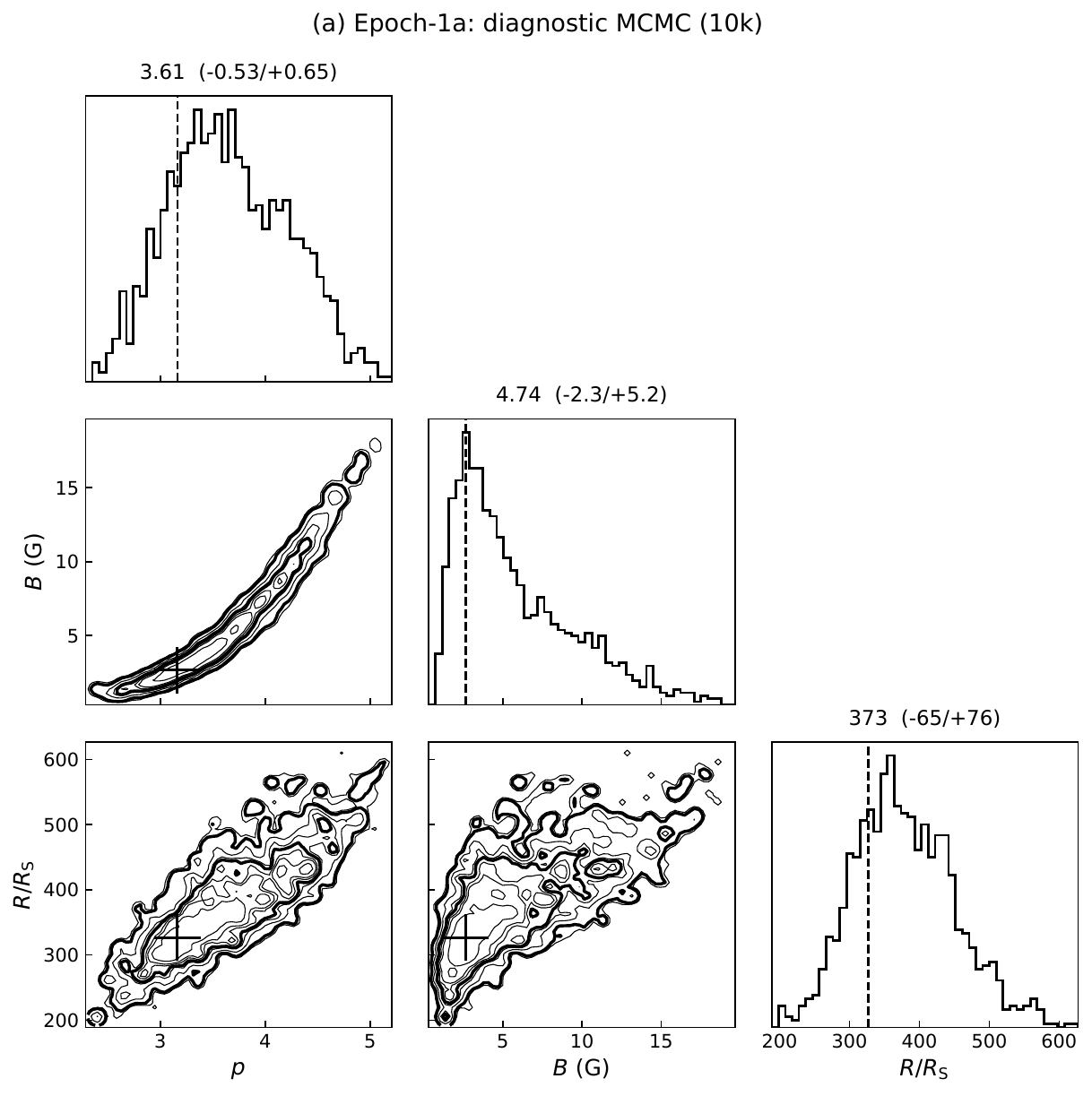}
  \end{minipage}
  \hfill
  \begin{minipage}[t]{0.49\textwidth}
    \centering
    \includegraphics[width=\linewidth]
    {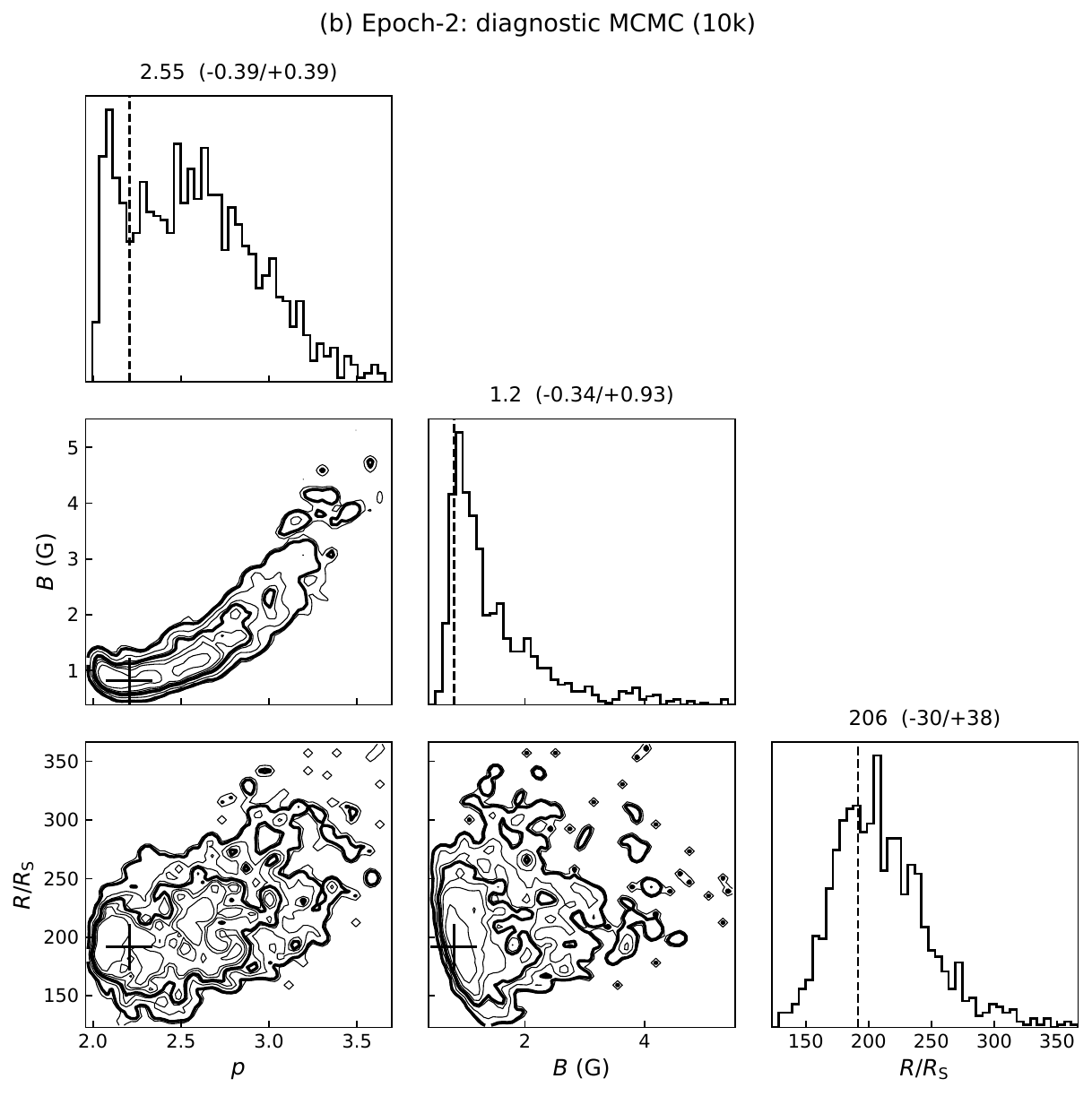}
  \end{minipage}
\end{center}
  \caption{
  Diagnostic MCMC distributions of $p$, $B$, and $R$ for Epoch-1a (left)
  and Epoch-2 (right).
  Diagonal and off-diagonal panels show one- and two-dimensional
  distributions, respectively.
  Thick contours enclose 68\% and 95\% of the retained samples; thinner
  curves show intermediate density levels.
  Plus signs and dashed vertical lines mark the least-squares best-fit values.
  }
  \label{fig:mcmc_corner}
\end{figure}

Figure~\ref{fig:mcmc_corner} shows the retained MCMC distributions of
$p$, $B$, and $R$ for Epoch-1a and Epoch-2.  Their strong, curved covariance
reflects the limited sampling of the synchrotron turnover region.
The least-squares best-fit points lie within the elongated distributions
but need not coincide with the one-dimensional peaks, illustrating why
$p$, $B$, and $R$ should not be interpreted independently.

\FloatBarrier
\bibliographystyle{aasjournalv7.1}

\end{document}